\documentclass[letterpaper,floatfix,reprint,aps,pre,longbibliography,superscriptaddress,nofootinbib]{revtex4-2} 

\usepackage{graphicx} 
\usepackage{silence}
\usepackage{gensymb}
\usepackage{tabularx}
\usepackage{color}
\usepackage{amsmath}
\usepackage{mathtools}
\usepackage{amssymb}
\usepackage{hyperref} 
\usepackage{xcolor}

\usepackage[version=3]{mhchem}
\usepackage[pass]{geometry}  

\begin{document}

\title{Optimal feedback control under stepwise equilibration and partial observation}

\author{Francesco Mottes}
\thanks{fmottes@seas.harvard.edu}
\affiliation{School of Engineering and Applied Sciences, Harvard University, Cambridge MA 02138}

\author{Michael P. Brenner}
\thanks{brenner@seas.harvard.edu}
\affiliation{School of Engineering and Applied Sciences, Harvard University, Cambridge MA 02138}


\begin{abstract}
Many microscopic machines rely on noisy signals to direct nonequilibrium transformations and operate in a timescale-separated regime. We consider feedback protocols in which a system equilibrates between rapid, measurement-conditioned changes of its energy landscape. In this limit, the partially observable control problem reduces exactly to a finite-horizon Bellman recursion over Hamiltonian updates. For a harmonic trap translated to a prescribed target under noisy position measurements, we solve this recursion analytically. The optimal protocol balances its response to the estimated fluctuation against progress toward the target, exploiting information early while enforcing the endpoint near the deadline. The minimum work decomposes into a positive thermodynamic-length transport cost and a negative information-enabled extraction term set by the fraction of equilibrium fluctuations resolved by the measurement. When a fixed intervention cost exceeds the asymptotic extraction per cycle, it selects a finite optimal number of cycles. We show that, in this translated harmonic case, the minimum thermodynamic cost of creating and resetting the measurement record always exceeds this threshold, making the optimal cycle count finite and the net work non-negative for any measurement channel.
\end{abstract}

\maketitle

\section{Introduction}

How to minimize the work required to drive a mesoscopic system through a finite-time transformation is a question of direct relevance to the study of both synthetic and biological molecular machines~\cite{seifert2012stochastic,tu2026nonequilibrium,cao2025stochastic}. A formulation commonly adopted in stochastic thermodynamics fixes the initial and terminal Hamiltonians and asks which finite-time protocol minimizes the expected work. In the absence of feedback, this problem has been extensively studied: open-loop protocols, fixed in advance rather than conditioned on observations, are characterized in slow-driving regimes by thermodynamic length~\cite{salamon1983thermodynamic,crooks2007measuring,sivak2012thermodynamic} and related optimal-control geometries~\cite{schmiedl2007optimal,rotskoff2017geometric,zhong2024beyond}. Finite sequences of discrete driven updates have also been studied~\cite{large2019optimal}.

The corresponding feedback problem, how to build optimal finite-horizon \textit{closed-loop} protocols, is less well understood. Information thermodynamics has established generalized second laws---upper bounds on measurement-enabled work extraction~\cite{sagawa2010generalized,horowitz2010nonequilibrium,parrondo2015thermodynamics}---but does not provide general constructive recipes. Exact optimal protocols have been derived in special cases: a single measurement~\cite{abreu2011extracting}, repeated feedback on passive and active particles~\cite{bauer2012efficiency,garcia2025optimal}, and noisy readouts exploited through Bayesian estimation~\cite{saha2022bayesian}.

In molecular machines and information engines, control actions are often triggered by signals that are only statistically correlated with the microscopic state. Mapping imperfect observations to conditional changes of an energy landscape, while accumulating work over a sequence of interventions, is naturally framed as a partially observable Markov decision process~\cite{bechhoefer2021control}.
Recent work has provided an exact solution for an overdamped particle in a harmonic trap with continuous relaxation between feedback cycles, partial observation, and costly measurements~\cite{panizon2026optimal}, followed by experimental validation~\cite{reinalter2026work}\footnote{Given the similarities to the analysis presented here, a more thorough comparison with these works is presented in Sec.~\ref{sec:discussion}.}.
Beyond the harmonic case, though, the optimization problem itself must usually be approximated with truncated belief states, approximate dynamic programming, or learned policies~\cite{engel2023optimal,whitelam2023demon}.

An alternative route to simplify the problem is to introduce a structural assumption of timescale separation between control events and local relaxation. We consider the case of discrete feedback systems that operate by an alternation of local equilibration in a fixed energy landscape and rapid conditioned switches between effective Hamiltonians~\cite{um2015total,ehrich2023energetic}. 
In this thermalization--observation--quench limit, the closed-loop thermodynamic design problem reduces exactly to a finite-horizon Bellman recursion over observation-conditioned Hamiltonian updates. Equilibration erases the memory of earlier cycles, so the only state the recursion propagates is the Hamiltonian itself. 

Many physical, chemical, and biomolecular processes naturally approximate these conditions~\cite{seifert2012stochastic,astumian2012microscopic}. In molecular motors such as F$_1$-ATPase~\cite{noji1997direct} and kinesin~\cite{svoboda1993direct}, internal chemomechanical signals gate sudden transitions between long-lived conformational states. In experimentally realized colloidal information engines, noisy position measurements trigger rapid shifts of an optical trap, after which the particle relaxes before the next cycle~\cite{toyabe2010experimental,paneru2018lossless}. The same separation of long-lived states connected by discrete switching events underlies effective stochastic kinetic schemes and reaction-network models~\cite{mugnai2020theoretical,hartich2021emergent}.

We solve a canonical continuous-state benchmark in which the recursion closes analytically: a one-dimensional harmonic trap with fixed stiffness, translated center, and noisy translation-covariant observation channel. In closed form, the optimal feedback law is a horizon-weighted compromise between the controller's estimate of the microstate and the prescribed target, exploiting the observed fluctuation early and enforcing the endpoint on the final step. We show that the minimum work splits into a positive thermodynamic-length transport cost---recovering the open-loop optimum---and a negative information term set by the fraction of equilibrium fluctuations resolved by the measurement. This decomposition identifies the number of feedback cycles at which the bare work balance crosses from dissipative driving to apparent fluctuation harvesting.

Charging a fixed physical cost per intervention then selects a finite optimal number of feedback cycles, provided the cost exceeds the asymptotic per-cycle extraction. Finally, we show that the minimum thermodynamic cost of creating and resetting the measurement record~\cite{landauer1961irreversibility} always exceeds this threshold in the translated-harmonic setting. The apparent harvest regime disappears, leaving the net work non-negative for any measurement channel and making the finite optimum thermodynamically unavoidable. These results are the analytically solvable instance of a more general structure: the same recursion governs any landscape and observation channel in the stepwise-equilibration limit, and can be solved numerically where analytic closure fails.

\begin{figure*}[t]
    \centering
    \includegraphics[width=\textwidth]{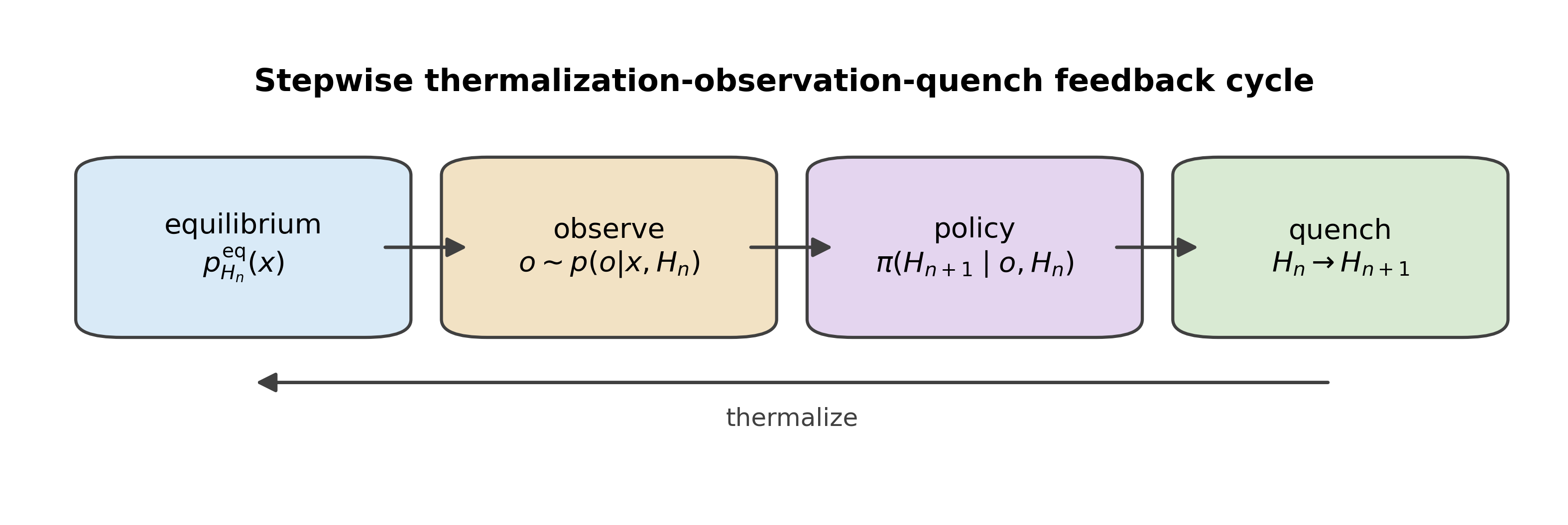}
    \caption{Stepwise thermalization--observation--quench feedback cycle. At the start of cycle $n$ the system is equilibrated under $H_n$, an observation $o\sim p(o\mid x,H_n)$ is acquired, and the policy $\pi(H_{n+1}\mid o,H_n)$ selects the next Hamiltonian. The resulting quench $H_n\to H_{n+1}$ is followed by re-thermalization under the new Hamiltonian before the next decision.}
    \label{fig:protocol-schematic}
\end{figure*}

\section{Stepwise-equilibration limit and Bellman recursion}

We consider protocols built from $N$ discrete feedback cycles indexed by $n=0,\dots,N-1$ and terminating at a prescribed target Hamiltonian $H_N=H_f$. Each cycle has three steps (Fig.~\ref{fig:protocol-schematic}): the system first relaxes under the current Hamiltonian $H_n$ to its equilibrium distribution, then an observation $o_n$ is acquired, and finally an observation-conditioned instantaneous quench updates the Hamiltonian to $H_{n+1}$. This structure arises naturally whenever local relaxation within a Hamiltonian basin is fast compared with the driven switching events that update it.

Two features of this regime simplify the optimal-control problem. First, equilibration collapses the state distribution at every decision time onto the equilibrium distribution of the current Hamiltonian
\begin{equation}\label{eq:equil}
p_{H_n}^{\mathrm{eq}}(x_n)=\frac{e^{-\beta H_n(x_n)}}{Z[H_n]},
\end{equation}
so earlier interventions leave no nonequilibrium memory. Second, work is performed only during the instantaneous quench. The subsequent relaxation exchanges heat with the bath but does no work, and the observation is assumed not to perturb the system or carry an explicit energetic cost---the case of costly measurements will be discussed in Section \ref{sec:info-landauer}. The controller can affect the expected work only through its observation-conditioned choice of the next Hamiltonian. The protocol is therefore controlled by a single decision rule, the policy $\pi(H_{n+1}\mid o_n,H_n)$, which maps the current Hamiltonian and observation to a distribution over post-quench Hamiltonians.

These assumptions give a simple product form for the likelihood of a full protocol realization. At each cycle, the microstate is sampled from the equilibrium distribution of the current Hamiltonian, the observation is drawn from the measurement kernel $p(o_n\mid x_n,H_n)$, and the next Hamiltonian is drawn from the policy $\pi$:
\begin{widetext}
\begin{equation}\label{eq:path-measure}
\mathbb P_{\pi}(x_{0:N-1},o_{0:N-1},H_{1:N}\mid H_0)
=
\prod_{n=0}^{N-1}
p_{H_n}^{\mathrm{eq}}(x_n)\,
p(o_n\mid x_n,H_n)\,
\pi(H_{n+1}\mid o_n,H_n),
\qquad H_N=H_f .
\end{equation}
\end{widetext}
This factorization is what makes the regime amenable to dynamic programming. Once $H_n$ is fixed, the current microstate and observation depend only on that Hamiltonian, and earlier cycles influence the future only through the Hamiltonian selected by the controller. The optimal-control problem therefore reduces to a sequential decision over the observation-conditioned Hamiltonian updates, with the Bellman recursion derived below.

\subsection{Bellman principle for minimum expected work}

We seek the minimum expected work needed to drive the system from the initial Hamiltonian $H_0$ to the prescribed target $H_f$ in $N$ feedback cycles. The construction has two parts: first identify the work cost of one cycle, then optimize the cumulative cost over the full horizon.

Because the quench is the only step that performs work, the stochastic work done \textit{on} the system in cycle $n$ is
\begin{equation}\label{eq:work-increment}
W_n = H_{n+1}(x_n)-H_n(x_n),
\end{equation}
where $W_n>0$ denotes work supplied by the controller and $W_n<0$ denotes work extracted during the feedback update. The protocol cost is the sum $\sum_{n=0}^{N-1}W_n$, minimized in expectation. Since the endpoints $H_0$ and $H_f$ are fixed, $\Delta F$ is constant, and minimizing the expected work is equivalent to minimizing the dissipation, i.e. the expected work in excess of the free-energy change. In the following we adopt the notation for the continuous-state form; the same expressions apply to discrete state, observation, or Hamiltonian spaces with integrals replaced by appropriate sums.

At cycle $n$ the controller does not see the microstate $x_n$ directly and observes only $o_n$. The observation marginal determines how often each signal is seen and can be derived from the equilibrium ensemble~\eqref{eq:equil} and the observation kernel:
\begin{equation}
p(o\mid H)=\int dx\,p(o\mid x,H)\,p_H^{\mathrm{eq}}(x).
\end{equation}
The corresponding microstate posterior is the controller's belief about $x$ after observing $o$, and follows from Bayes' rule:
\begin{equation}
p(x\mid o,H)=\frac{p(o\mid x,H)\,p_H^{\mathrm{eq}}(x)}{\int dx'\,p(o\mid x',H)\,p_H^{\mathrm{eq}}(x')}.
\end{equation}

Given the current Hamiltonian $H$ and observation $o$, a policy $\pi(H'\mid o,H)$ selects a post-quench Hamiltonian $H'$. The system is then allowed to relax to equilibrium under $H'$ before the next cycle begins. Averaging the single-cycle work~\eqref{eq:work-increment} over the posterior gives the expected immediate work of choosing $H'$ after observing $o$,
\begin{equation}
\begin{aligned}
w(H,o;H')&=\mathbb{E}\!\left[W_n\,\middle|\,H,o;H'\right]\\
&=\int dx\,p(x\mid o,H)\bigl(H'(x)-H(x)\bigr).
\end{aligned}
\end{equation}

This is only the cost of the current cycle. Minimizing it greedily is generally suboptimal: a cheap quench now may leave the system in a Hamiltonian from which reaching $H_f$ is expensive later. 
To express the trade-off, suppose that after the current cycle the controller continues optimally. Let the $r$-step value function $V_r(H)$ denote the minimum expected remaining work needed to reach $H_f$ in $r$ additional feedback steps, starting from equilibrium under Hamiltonian $H$ (so at cycle $n$ of an $N$-step protocol, $r=N-n$).
Additivity of work and the factorized path measure~\eqref{eq:path-measure} give the expected cost of using a current-cycle policy $\pi$ with $r$ steps remaining:
\begin{widetext}
\begin{equation}\label{eq:policy-expectation}
J_r^{\pi}(H)=\int do\,p(o\mid H)\int dH'\,\pi(H'\mid o,H)\bigl[w(H,o;H')+V_{r-1}(H')\bigr].
\end{equation}
\end{widetext}
For a fixed observation $o$, the policy $\pi(\cdot\mid o,H)$ only forms a weighted average of the costs associated with different choices of $H'$. An average cannot be lower than its smallest term, so randomizing cannot improve on placing all probability on a minimizing post-quench Hamiltonian. Therefore, an optimal deterministic policy always exists~\cite{bertsekas2012dynamic}. Optimizing the current-cycle policy gives the Bellman recursion for the value function $V_r(H)=\inf_\pi J_r^{\pi}(H)$:
\begin{widetext}
\begin{equation}\label{eq:bellman}
V_r(H)=
\mathbb{E}_{o\sim p(o\mid H)}
\left[
\min_{H'}\left(w(H,o;H')+V_{r-1}(H')\right)
\right],
\qquad
V_0(H)=
\begin{cases}
0,& H=H_f,\\
+\infty,& H\neq H_f.
\end{cases}
\end{equation}
\end{widetext}
The minimization balances the posterior-averaged work cost, or gain, of the current quench against the optimal continuation cost from the Hamiltonian chosen by that quench. The terminal condition enforces the prescribed endpoint by assigning infinite cost to protocols that end away from $H_f$. Thus optimal feedback design in the thermalize--observe--quench limit becomes an exact finite-horizon sequential decision problem over observation-conditioned Hamiltonian updates. In control-theoretic language, Eq.~\eqref{eq:bellman} is the belief-state Bellman equation for a partially observed Markov decision process~\cite{aastrom1965optimal,bertsekas2012dynamic} with the stochastic-work cost of Eq.~\eqref{eq:work-increment}.

This formulation also clarifies limiting cases and extensions. Full observation is recovered by setting $o=x$: the posterior collapses to a delta function, and the recursion reduces to a pointwise minimization at the observed microstate. Incomplete relaxation would enlarge the controlled state to include the nonequilibrium information carried between interventions. Restricting the Hamiltonian to a parametric family instead converts the minimization over $H'$ into a finite-dimensional control problem, mirroring the limited-control setting in which exact open-loop protocols have been obtained arbitrarily far from equilibrium~\cite{zhong2022limited}. When analytic closure is not available, this structure also identifies the state--action decomposition needed for computational approaches.

\section{Translated harmonic trap under partial observation}

We now analyze the simplest continuous-state model that admits a closed-form solution, with an explicit optimal protocol and a cost that separates into transparent physical contributions. Consider the one-dimensional harmonic family
\begin{equation}\label{eq:harmonic}
H_\lambda(x)=\frac{k}{2}(x-\lambda)^2,
\end{equation}
with fixed stiffness $k$ and prescribed displacement from $\lambda=0$ to $\lambda=L$ in $N$ feedback steps. Within this family, the control variable is the post-quench trap center $\lambda_{n+1}$. 

To evaluate the recursion, we first need to compute the average per-cycle work. Because this family is labeled by the trap center alone, we write $\lambda$ for $H_\lambda$ in function arguments throughout. Let $\hat{x}(o;\lambda):=\mathbb E[x\mid o,\lambda]$ be the controller's best estimate of the microstate given the observation---the mean of the (posterior) belief $p(x\mid o,\lambda)$ introduced above. 
At equilibrium, the microstate fluctuates about the trap center with variance $\sigma_{\mathrm{eq}}^2=k_{\mathrm B}T/k$. By the law of total variance this fluctuation splits into two parts: the variability of the estimated state $\hat{x}$ as the measurement $o$ changes (resolved by the detector, denoted by $\eta^2$) and the average uncertainty on the actual underlying state $x$ given the observation $o$,
\begin{equation}\label{eq:total-variance-split}
\sigma_{\mathrm{eq}}^2
=\underbrace{\operatorname{Var}_{o\mid\lambda}\!\bigl(\hat{x}(o;\lambda)\bigr)}_{\eta^2\ \text{(resolved)}}
+ \underbrace{\mathbb E_{o\mid\lambda}\!\bigl[\operatorname{Var}(x\mid o,\lambda)\bigr]}_{\text{unresolved}} .
\end{equation}

In the fixed-stiffness translated harmonic family, only the resolved part survives into the work. The per-cycle work is the posterior-averaged energy change of the quench, $w(\lambda,o;\lambda')=\tfrac{k}{2}\,\mathbb E\!\left[(x-\lambda')^2-(x-\lambda)^2\mid o,\lambda\right]$, and the identity $\mathbb E[(x-a)^2\mid o,\lambda]=\operatorname{Var}(x\mid o,\lambda)+(\hat{x}-a)^2$ gives each squared displacement the same unresolved scatter $\operatorname{Var}(x\mid o,\lambda)$ plus a term in the estimate alone. Subtracting the two displacements removes the unresolved part, leaving a work that depends on the estimate only,
\begin{equation}\label{eq:variance-cancellation}
w(\lambda,o;\lambda')=\frac{k}{2}(\hat{x}-\lambda')^2-\frac{k}{2}(\hat{x}-\lambda)^2.
\end{equation}
The partial-observation Bellman equation therefore takes its full-observation form with the microstate $x$ replaced by its estimate $\hat{x}(o;\lambda)$, now a sufficient statistic for the whole control problem---the belief state of the previous section collapses to its mean.

The quality of the detector is naturally measured by its resolved fraction of the total fluctuations,
\begin{equation}
\chi:=\frac{\eta^2}{\sigma_{\mathrm{eq}}^2}\in[0,1].
\end{equation}
The resolved-variance fraction $\chi$ interpolates between a useless detector ($\chi=0$), whose estimate carries no information about $x$, and perfect state observation ($\chi=1$).

We can now solve the recursion by exploiting its quadratic structure. The stage cost~\eqref{eq:variance-cancellation} is quadratic in the trap centers and the terminal condition fixes the single point $\lambda=L$, so each Bellman step carries a quadratic value function to another quadratic centered on the target. We therefore try the ansatz
\begin{equation}\label{eq:quadratic-ansatz}
V_r(\lambda)=a_r(\lambda-L)^2+b_r .
\end{equation}
Substituting into Eq.~\eqref{eq:bellman} transforms the functional recursion into a pair of separate scalar recursions for the curvature $a_r$ and the offset $b_r$,
\begin{equation}\label{eq:anbn-recursion}
a_r=\frac{k\,a_{r-1}}{k+2a_{r-1}},\qquad b_r=b_{r-1}-\frac{k^2\eta^2}{2(k+2a_{r-1})},
\end{equation}
with terminal conditions $a_0=+\infty$, $b_0=0$ that enforce $V_0(\lambda)=0$ at $\lambda=L$ and $+\infty$ otherwise.

This reduction rests on the crucial assumption that the resolved variance $\eta^2$ is the same at every trap center. This is the case when the detector reads the microstate's position relative to the trap center rather than an absolute lab-frame coordinate. 
The translation covariance property can be safely assumed for a broad class of detectors commonly adopted in feedback-trap experiments, such as additive-noise readouts $o=x+\xi$ and relative thresholds or photon counts. A lab-frame detector whose informativeness varied with absolute position would instead make $\eta^2=\eta^2(\lambda)$, re-couple the two recursions, and prevent a closed form solution. A more precise treatment and measurement channel examples are described in Supplementary Section~\ref{sec:supp-translation-covariance}.

The scalar recursions for the value function coefficients then solve in closed form,
\begin{equation}\label{eq:anbn-closed}
a_r=\frac{k}{2r},\qquad b_r=-\frac{k\eta^2}{2}\bigl(r-\mathcal{H}_r\bigr),
\end{equation}
with $\mathcal{H}_r=\sum_{q=1}^r q^{-1}$ the $r$th harmonic number. The intermediate algebra is in Supplementary Section~\ref{sec:supp-pomdp}.

\begin{figure}[t]
\centering
\includegraphics[width=\columnwidth]{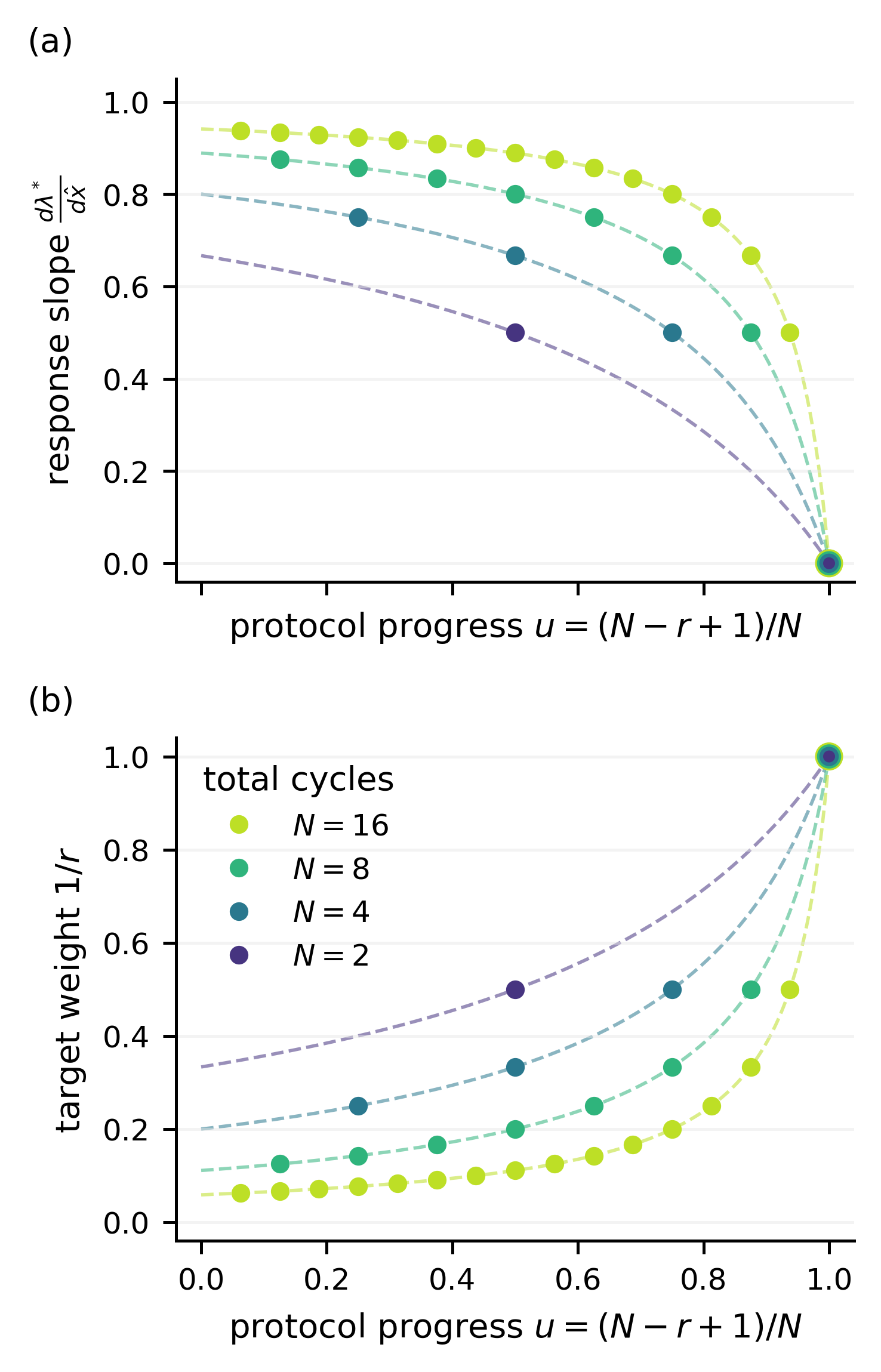}
\caption{Time dependence of the optimal feedback law $\lambda^*(o;r)=\tfrac{r-1}{r}\hat{x}(o;\lambda)+\tfrac{1}{r}L$ for total horizons $N\in\{2,4,8,16\}$, plotted against normalized decision number $u=(N-r+1)/N$. (a) The estimate-response slope $d\lambda^*/d\hat{x}=(r-1)/r$ decreases to zero as the terminal constraint takes over. (b) The target weight $1/r$ increases toward unity at the final step.}
\label{fig:feedback-law}
\end{figure}

The inner minimization over the post-quench center yields the optimal feedback law,
\begin{equation}\label{eq:feedback-law}
\lambda^*(o;r)=\frac{r-1}{r}\,\hat{x}(o;\lambda)+\frac{1}{r}\,L.
\end{equation}
The optimal trap center is a weighted average of the estimate and the target, balancing work extraction against the terminal constraint. When many cycles remain ($r$ large), the weight $(r-1)/r$ places the trap almost entirely on the estimate $\hat{x}$, harvesting the work available from the next quench (Fig.~\ref{fig:feedback-law}(a)). As fewer cycles remain (small $r$), the growing weight $1/r$ shifts the trap toward the target $L$, until at the final cycle ($r=1$) it sits entirely on $L$ and the protocol reaches the prescribed final Hamiltonian (Fig.~\ref{fig:feedback-law}(b)). The handoff from tracking to targeting is gradual and parameter-free in the remaining horizon $r$, independent of the detector quality $\chi$. In the many-cycle limit $r\to\infty$ the terminal pull vanishes and the law reduces to pure tracking of the microstate estimate, recovering the steady-state center-feedback policy of cyclically operated Brownian information machines~\cite{bauer2012efficiency}.

Evaluating the value function at the initial horizon $r=N$ and the initial trap position $\lambda=0$ gives the minimum expected work over the whole protocol:
\begin{equation}\label{eq:min-work}
W_N^*=
\frac{kL^2}{2N}-\frac{k_{\mathrm B}T}{2}\,\chi\,\left(N-\mathcal{H}_N\right).
\end{equation}
The two terms decouple cleanly: the first depends only on the trap stiffness and target displacement and is independent of the detector, while the second depends only on the detector quality $\chi$ and the number of cycles, not on the target displacement. The fully observed problem is recovered as the special case $\chi=1$, for which the estimate resolves the full equilibrium fluctuation and $\hat{x}(o;\lambda)=x$.


\subsection{Transport--information decomposition of the minimum work}

The first term in Eq.~\eqref{eq:min-work} has a purely geometric meaning: it is the work of transporting the equilibrium distribution from $\lambda=0$ to $\lambda=L$. To show this, we compute the Fisher metric of the equilibrium family and integrate it to find the thermodynamic length separating the two endpoints. Because the partition function $Z_\lambda$ is independent of the trap center, a shift of the center changes the equilibrium log-density at the rate $\partial_\lambda \ln p_\lambda^{\mathrm{eq}}(x)=\beta k(x-\lambda)$, and the variance of this rate over the equilibrium ensemble defines the Fisher metric,
\begin{equation}\label{eq:fisher}
g_{\lambda\lambda}=\mathbb E\!\left[\bigl(\partial_\lambda \ln p_\lambda^{\mathrm{eq}}\bigr)^2\right]=\beta^2 k^2\sigma_{\mathrm{eq}}^2=\beta k.
\end{equation}
The metric measures how distinguishable neighboring members of the family are, fixing the local thermodynamic cost of displacing the trap~\cite{ito2018stochastic}.
Here it is constant along the family, so the equilibrium states are uniformly distinguishable and the trap moves through them at a fixed cost per unit displacement. Integrating $\sqrt{g_{\lambda\lambda}}$ accumulates this cost into the thermodynamic length, the geometric distance between the endpoints in probability space:
\begin{equation}\label{eq:length}
\mathcal{L}=\int_0^L\!\sqrt{g_{\lambda\lambda}}\,d\lambda=\sqrt{\beta k}\,L .
\end{equation}

Substituting $\mathcal{L}^2=\beta kL^2$ into Eq.~\eqref{eq:min-work} recasts the minimum work as
\begin{widetext}
\begin{equation}\label{eq:decomposition}
W_N^*=
\underbrace{\frac{k_\mathrm{B}T}{2}\frac{\mathcal{L}^2}{N}}_{\text{thermodynamic-length transport cost}}
-\underbrace{\frac{k_\mathrm{B}T}{2}\,\chi\,\left(N-\mathcal{H}_N\right)}_{\text{observation-enabled extraction}}.
\end{equation}
\end{widetext}
The first term is the finite-horizon cost of transporting the equilibrium distribution across a thermodynamic distance $\mathcal{L}$: it is the classical minimum dissipation of a quasistatic process split into $N$ equilibrated steps, $k_{\mathrm B}T\,\mathcal{L}^2/(2N)$~\cite{nulton1985quasistatic,sivak2012thermodynamic}. The second is the cumulative gain made possible by the information the observation channel resolves, and is the genuinely closed-loop contribution. The harmonic example combines both contributions in a single exact closed-form expression.

\begin{figure}[t]
\centering
\includegraphics[width=\columnwidth]{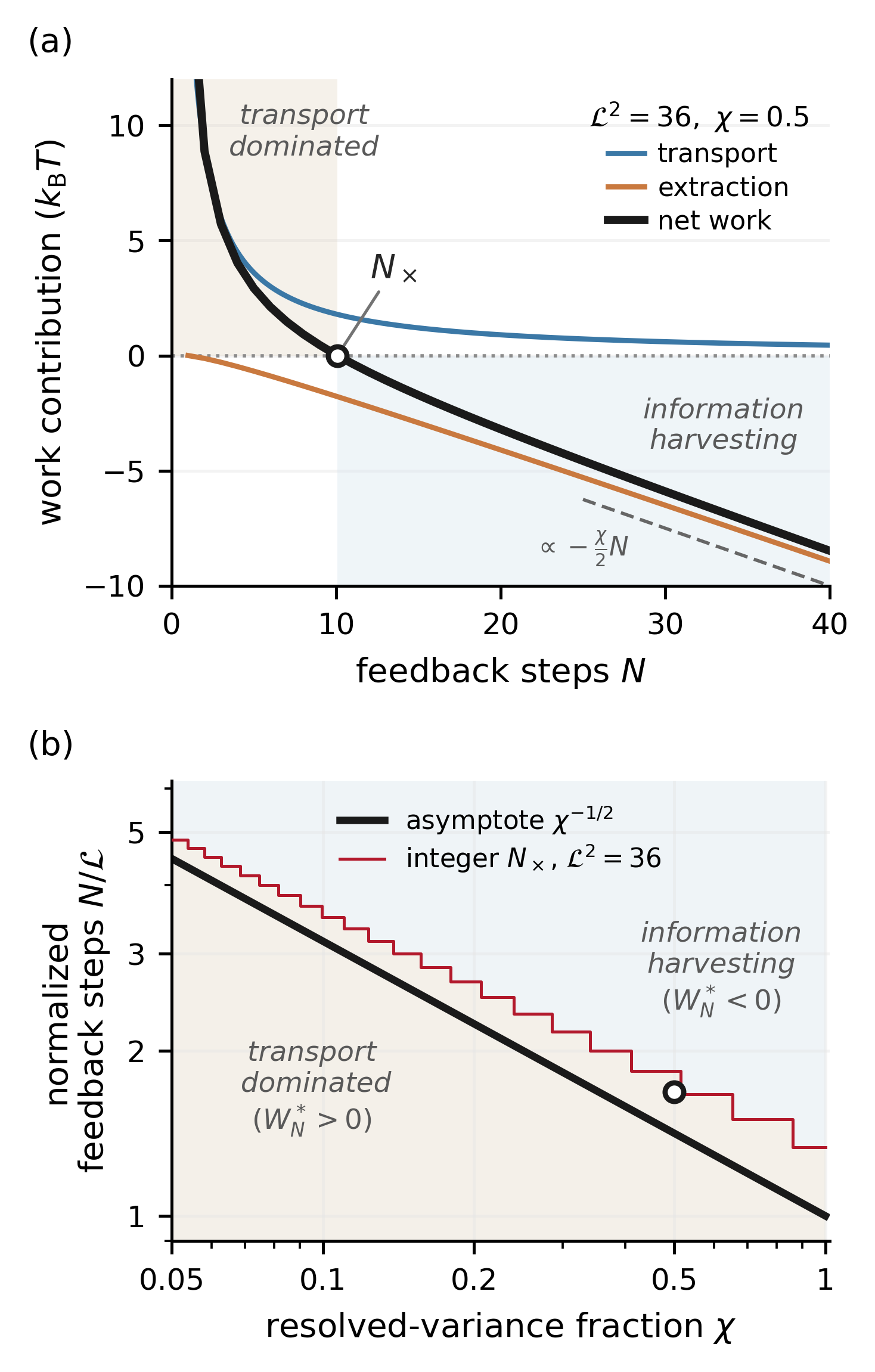}
\caption{Transport--information decomposition of the optimal work. (a) Positive transport contribution $\mathcal{L}^2/(2N)$, negative observation-enabled contribution $-\tfrac{1}{2}\chi(N-\mathcal{H}_N)$, and net work $W_N^*/(k_{\mathrm B}T)$ for $\mathcal{L}^2=36$ and $\chi=0.5$; the zero crossing $N_\times$ separates transport-dominated protocols from bare-work information harvesting. (b) Smallest integer $N_\times$ with $W_N^*\leq0$, normalized by $\mathcal{L}$, compared with the large-distance asymptote $\chi^{-1/2}$.}
\label{fig:work-decomposition}
\end{figure}

The decomposition has an immediate zero crossing: Eq.~\eqref{eq:decomposition} changes sign at $N_\times$ feedback cycles, where $\mathcal{L}^2=\chi N_\times(N_\times-\mathcal{H}_{N_\times})$.
Below $N_\times$ the transport cost dominates and the protocol dissipates work. Above $N_\times$ the observation-enabled extraction exceeds the transport cost, and the protocol becomes a net harvester of work (Fig.~\ref{fig:work-decomposition}(a)). This harvest is bare: it counts $W_N^*$ alone and does not yet debit the thermodynamic cost of creating and resetting the measurement record, which we account for in Sec.~\ref{sec:info-landauer}.
Dropping the $\mathcal{O}(\ln N_\times)$ harmonic-number correction gives the closed-form asymptote shown in Fig.~\ref{fig:work-decomposition}(b),
\begin{equation}\label{eq:Ncross}
N_\times\;\simeq\;\frac{\mathcal{L}}{\sqrt{\chi}}\;=\;\frac{L}{\sigma_{\mathrm{eq}}\sqrt{\chi}}.
\end{equation}

The crossing requires no fitted parameters: once $\chi$ is calibrated from the detector's signal-to-noise and $L,k$ are set by the target displacement and trap stiffness, $N_\times$ follows directly, which makes Eq.~\eqref{eq:Ncross} testable in translated-trap feedback engines of the kind realized in Refs.~\cite{toyabe2010experimental,paneru2018lossless,saha2022bayesian}. A crossover of exactly this kind---from dissipative dragging to negative average work as the number of feedback updates grows---has recently been observed in a fixed-stiffness translated optical trap, with feedback policies learned in situ by reinforcement learning~\cite{reinalter2026work}. A moderate displacement $L=6\,\sigma_{\mathrm{eq}}$ ($\mathcal{L}^2=36$, the value plotted in Fig.~\ref{fig:work-decomposition}) with a detector resolving half the equilibrium variance ($\chi=0.5$) places the exact sign change at $N_\times=11$ cycles, well within the range of existing protocols.

The two sides of the zero crossing correspond to physically distinct regimes. For $N\ll N_\times$ the dissipation is geometric and detector-independent,
\begin{equation}\label{eq:regime-transport}
W_N^*\;\sim\;\frac{k_{\mathrm B}T\,\mathcal{L}^2}{2N}
\qquad(N\ll N_\times),
\end{equation}
and the protocol pays the full Sivak--Crooks-like thermodynamic-length cost of the transport. For $N\gg N_\times$ the transport term is subleading and the harvest rate saturates at $\chi\,k_{\mathrm B}T/2$ per cycle,
\begin{equation}\label{eq:regime-info}
W_N^*\;\sim\;-\frac{k_{\mathrm B}T}{2}\,\chi\,N
\qquad(N\gg N_\times),
\end{equation}
independently of the target displacement $L$. The decomposition~\eqref{eq:decomposition} therefore exposes a transport--information crossover structure whose scale is set by $\mathcal{L}$ and~$\chi$.

\subsection{Optimal number of feedback cycles under per-cycle overhead}
\label{sec:cadence}

Each driven intervention carries a cost---measurement, actuation, or fuel---that $W_N^*$ does not account for. With a fixed overhead $c$ per cycle, we must minimize not just the work but the whole accounted total
\begin{equation}\label{eq:total-cost}
\mathcal{J}_N=W_N^*+cN.
\end{equation}
The optimal number of feedback cycles,
\begin{equation}\label{eq:optimal-N}
N^*=\operatorname*{argmin}_N \mathcal{J}_N ,
\end{equation}
balances the transport cost, the observation-enabled extraction, and the overhead. The marginal-balance condition $\Delta \mathcal{J}_N:=\mathcal{J}_{N+1}-\mathcal{J}_N=0$ reduces exactly to a quadratic in $N$ (Supplementary Section~\ref{sec:supp-cadence}):
\begin{equation}\label{eq:cadence-quadratic}
(\hat{c}-\chi)\,N^2+\hat{c}\,N-\mathcal{L}^2=0,
\end{equation}
with the dimensionless overhead defined as $\hat{c}=2c/(k_{\mathrm B}T)$. The unique positive root is
\begin{equation}\label{eq:Nstar}
N^*=\frac{-\hat{c}+\sqrt{\hat{c}^{\,2}+4(\hat{c}-\chi)\,\mathcal{L}^2}}{2(\hat{c}-\chi)},\qquad \hat{c}>\chi.
\end{equation}
Equation~\eqref{eq:Nstar} is the real-valued root of the marginal-balance condition. For the integer protocol the optimal number of feedback cycles is $\lceil N^*\rceil$, the smallest integer at which the marginal cost is non-negative. A finite optimum exists when the overhead meets or exceeds the asymptotic per-cycle extraction, $\hat{c}\geq\chi$. Below that threshold each added cycle extracts more work than it costs and arbitrarily many feedback cycles are optimal. Whether a physical overhead can ever fall below this threshold is a thermodynamic question, which we settle in Sec.~\ref{sec:info-landauer}. In the absence of useful observation ($\chi=0$), Eq.~\eqref{eq:Nstar} reduces to $N^*\approx\mathcal{L}/\sqrt{\hat{c}}$ at large thermodynamic distance, recovering the classical open-loop transport--overhead balance. The harmonic benchmark thus yields an explicit design rule for the number of feedback cycles under imperfect measurement.

\subsection{Landauer cost and feedback feasibility}
\label{sec:info-landauer}

The preceding optimization over the number of feedback cycles treated the per-cycle overhead $c$ as a phenomenological cost, meant to represent measurement, actuation, fuel, or other resources not included in $W_N^*$. It is therefore natural to ask whether the condition for a finite optimum, $\hat c>\chi$, is optional or forced by thermodynamics. We show that it is forced once the controller's memory is included in the accounting. Each cycle creates a measurement record correlated with the microstate. We assume that the controller's memory is returned to a standard state after every cycle. The minimum combined thermodynamic cost of creating and resetting this correlated record is at least $k_{\mathrm B}T\,I(x;o)$ per cycle~\cite{landauer1961irreversibility,parrondo2015thermodynamics}, where $I(x;o)$ is the mutual information between the true microstate $x$ and the observation $o$; this is the information-theoretic form of Landauer's bound, to which it reduces for a single erased bit. Because the memory is reset after every cycle, this lower bound applies separately at each step and accumulates to $N\,k_{\mathrm B}T\,I(x;o)$ over the protocol~\cite{horowitz2010nonequilibrium}. This gives a thermodynamic floor on the overhead $c$; for a Gaussian readout, that floor lies above the threshold for a finite optimum and excludes the formal regime in which arbitrarily many feedback cycles would extract unbounded work.

To make this statement explicit we must price the information stored by the controller. The closed-form work result depends on the observation channel only through the resolved-variance fraction $\chi$, whereas the information cost depends in general on the full channel statistics. We evaluate it for the additive Gaussian channel $o=x+\xi$ with $\xi\sim\mathcal{N}(0,\Delta^2)$, the standard detector model. Because in this case the equilibrium distribution is itself Gaussian, this channel is also the extremal one: at a fixed resolved fraction $\chi$, it minimizes the mutual information among all observation channels, by the rate--distortion theorem for a Gaussian source~\cite{cover1999elements}. Therefore, the floor it sets bounds every alternative channel model from below. Explicitly,
\begin{equation}\label{eq:landauer-mutual-information}
I(x;o)=\tfrac{1}{2}\ln\frac{\sigma_{\mathrm{eq}}^2+\Delta^2}{\Delta^2}=\tfrac{1}{2}\ln\frac{1}{1-\chi},
\end{equation}
giving $I(x;o)\geq\tfrac{1}{2}\ln[1/(1-\chi)]$ for any observation channel and the channel-independent overhead floor
\begin{equation}\label{eq:landauer-floor}
\hat c\geq \hat c_{\mathrm L}:=\ln\frac{1}{1-\chi}.
\end{equation}
The preceding optimization over the number of feedback cycles rests on no special property of the noise model, only on the Gaussian equilibrium of the harmonic family.

\begin{figure}[t]
\centering
\includegraphics[width=\columnwidth]{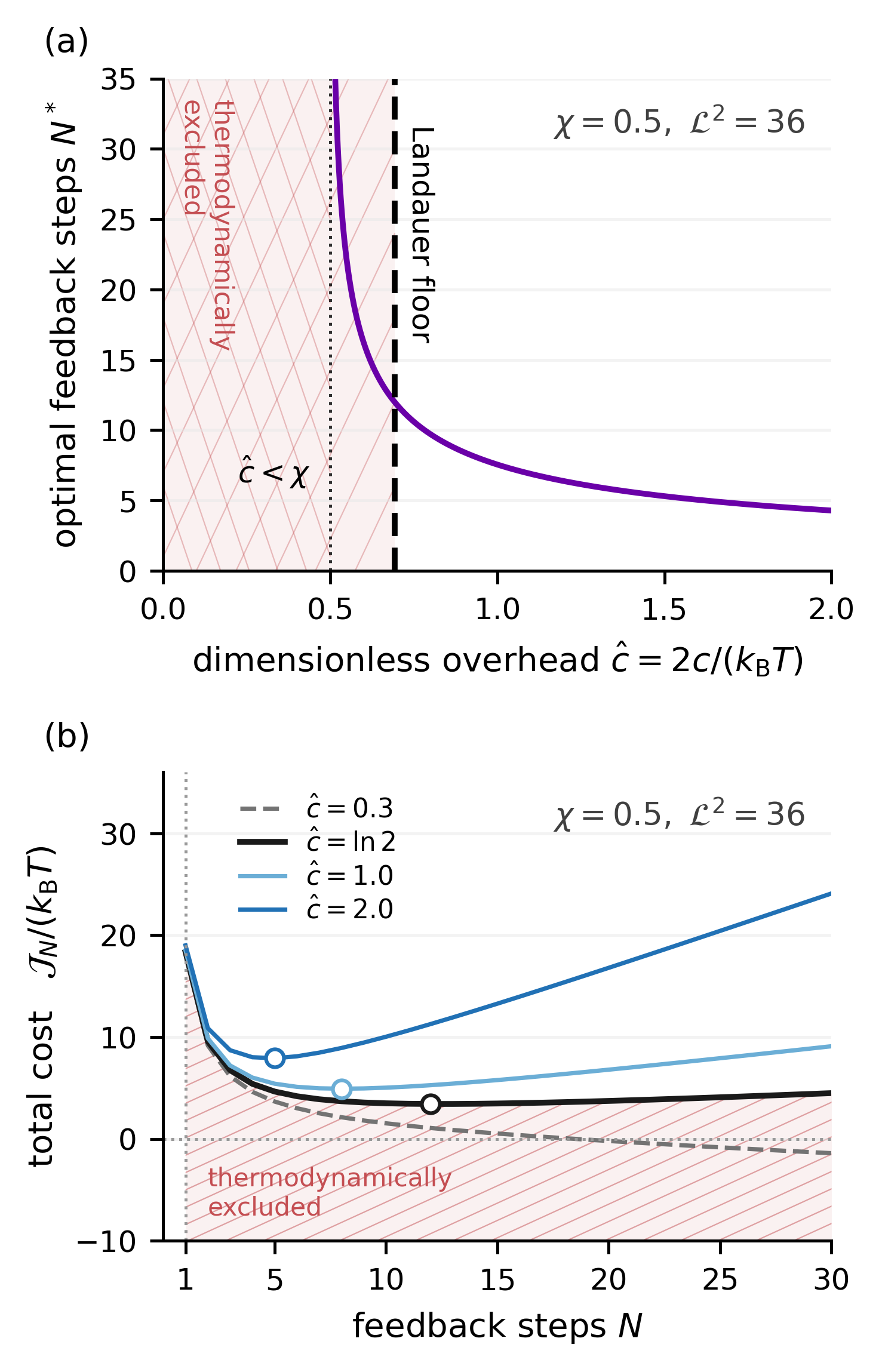}
\caption{Optimal number of feedback cycles selected by a per-cycle overhead for $\chi=0.5$ and $\mathcal{L}^2=36$. (a) Positive root $N^*$ of the marginal-balance condition as a function of dimensionless overhead $\hat c=2c/(k_{\mathrm B}T)$; the dotted line marks the formal threshold $\hat c=\chi$, while the dashed line marks the Landauer floor $\hat c=\ln[1/(1-\chi)]=\ln2$. (b) Total cost $\mathcal{J}_N/(k_{\mathrm B}T)$ vs.\ number of feedback cycles for several overheads, with open circles marking finite minima when $\hat c>\chi$. Shaded regions indicate overheads or costs that are thermodynamically excluded once the cost of creating and resetting the measurement record is debited.}
\label{fig:optimal-n-cycles}
\end{figure}

The significance of this floor is clearest for the translated harmonic family, where the partition function is independent of the trap center and hence $\Delta F=0$. The feedback-generalized second law of Sagawa and Ueda~\cite{sagawa2010generalized} bounds the extractable work per cycle by $k_{\mathrm B}T\,I(x;o)$. Our protocol achieves the asymptotic extraction $\chi\,k_{\mathrm B}T/2$ per cycle, while the Gaussian-channel bound is $(k_{\mathrm B}T/2)\ln[1/(1-\chi)]$. Since $\ln[1/(1-\chi)]>\chi$ for every $\chi>0$, the Landauer floor lies above the finite-optimum threshold $\hat c=\chi$ (Fig.~\ref{fig:optimal-n-cycles}(a)). Thus the finite optimum of Sec.~\ref{sec:cadence} is not merely a modeling assumption: once the cost of creating and resetting the measurement record is debited, the sub-threshold branch with arbitrarily many feedback cycles is thermodynamically inaccessible.

The same accounting also excludes the apparent unlimited harvest regime. If the overhead is set at its Landauer floor, $c=k_{\mathrm B}T\,I(x;o)$, the total accounted work is
\begin{equation}\label{eq:total-net-balance}
\begin{aligned}
W_N^{\mathrm{tot}}&:=W_N^*+N\,k_{\mathrm B}T\,I(x;o)\\
&=\frac{k_{\mathrm B}T\,\mathcal{L}^2}{2N}+\frac{k_{\mathrm B}T\,N}{2}\bigl[\ln\tfrac{1}{1-\chi}-\chi\bigr]+\frac{k_{\mathrm B}T\,\chi}{2}\,\mathcal{H}_N,
\end{aligned}
\end{equation}
which is a sum of three non-negative terms. Hence $W_N^{\mathrm{tot}}\geq0$ throughout the $(\chi,N)$ plane, and the negative-work regime of $W_N^*$ is revealed as a bookkeeping artifact of pricing observations at zero rather than a source of net work after the measurement record is reset (Fig.~\ref{fig:optimal-n-cycles}(b)).

What survives is therefore not free measurement-enabled work, but a physical design rule for the number of feedback cycles. Additional cycles reduce the transport term $k_{\mathrm B}T\mathcal{L}^2/(2N)$, while imperfect measurements add a record-creation and reset cost whose size is set by the channel. The gap to the information-theoretic ceiling is also informative: the asymptotic extraction is a fraction $r(\chi)=\chi/\ln[1/(1-\chi)]$ of the Sagawa--Ueda bound, approaching unity for weak measurements and falling toward zero for strong measurements, where translation-only control cannot exploit all of the acquired information---consistent with the known result that a Brownian information machine actuating only the trap center under-performs one that also controls the stiffness~\cite{bauer2012efficiency}. The full Gaussian-channel derivation and saturation comparison are given in Supplementary Section~\ref{sec:supp-gaussian}.

\section{Discussion}
\label{sec:discussion}

Our results rely on a structural physical assumption of separation of timescales. When a system re-equilibrates between measurement-conditioned quenches, the memory of earlier interventions is erased and optimal feedback design---generically an intractable partially observed control problem---becomes an exact finite-horizon Bellman recursion whose only propagated state is the Hamiltonian itself. On the translated harmonic family the recursion closes analytically. The optimal feedback law is a horizon-weighted compromise, tracking the estimated fluctuation when many cycles remain and pinning the trap to the target as the deadline nears. The minimum work it achieves splits exactly into a geometric transport cost---the open-loop optimum---and an information credit set by the resolved fraction $\chi$ of the equilibrium fluctuations.
Their crossing sets the number of feedback cycles $N_\times$ beyond which the protocol appears to harvest work, and their balance against a per-cycle overhead selects a finite optimal number of cycles. The analysis of Sec.~\ref{sec:info-landauer} closes the accounting: once the cost of creating and resetting the measurement record is included at its thermodynamic lower bound, the apparent harvest regime disappears, leaving a design rule, Eq.~\eqref{eq:Nstar}, for how many feedback cycles are worth performing. Both predictions require no fitted parameters once the trap and the detector are calibrated, placing a direct test of the framework within reach of existing feedback-trap experiments~\cite{toyabe2010experimental,paneru2018lossless,saha2022bayesian,reinalter2026work}.

These results are naturally read alongside the recent work of Panizon~\cite{panizon2026optimal}, who solves the same physical benchmark---an overdamped particle in a fixed-stiffness harmonic trap, steered between prescribed centers by measurement-conditioned displacements---as a discrete-time POMDP with costly measurements, a solution validated experimentally~\cite{reinalter2026work}. In our stepwise-equilibration limit, their belief update collapses onto the equilibrium distribution of the current trap and their optimal trap placement reduces to the horizon-weighted compromise of Eq.~\eqref{eq:feedback-law}. The correspondence extends to the energetics: in their engine, a fully resolving measurement never pays once its cost exceeds $k_{\mathrm B}T/2$---our threshold $c>\chi\,k_{\mathrm B}T/2$ for a finite optimum at perfect resolution---and measurement ceases near the deadline, where information can no longer repay its cost. Where their measurement cost is a phenomenological device parameter, however, we bound the overhead from below by the minimum cost of creating and resetting the measurement record, ensuring that the optimal protocol uses finitely many feedback cycles for any observation channel.
The two works generalize along different axes. Panizon retains the finite relaxation kinetics and treats the measurement itself as a control variable, decoupled from the driving by the same linear-quadratic-Gaussian structure that confines the analysis to the harmonic setting. We take the opposite trade: complete equilibration removes the kinetics, so that time enters only through the number of interventions, but the recursion then holds for arbitrary energy landscapes, state spaces, and observation channels. 

The assumptions that make this exactness possible also delimit it. The separation of timescales is the strongest: whenever driving competes with local relaxation, or the landscape supports long-lived metastable states, equilibration between interventions is incomplete, memory persists across cycles, and the recursion must propagate a genuine belief state rather than the Hamiltonian alone. The framework likewise counts interventions rather than elapsed time, so the kinetic cost of finite-rate driving enters only through the overhead $c$. Finally, no closed form should be expected beyond the fixed-stiffness harmonic family and translation-covariant channels. The recursion itself, however, remains exact and can be iterated numerically, complementing recent computational approaches to protocol design based on learned policies~\cite{engel2023optimal,whitelam2023demon} and gradient-based optimization of kinetic models~\cite{mottes2026gradient}.

The same boundaries mark some natural next steps. The recursion can be used to search for optimal feedback protocols on anharmonic or discrete energy landscapes, with the kinetic schemes of molecular machines as obvious candidates. A second direction is to extend the analysis to multidimensional systems and multiple coupled control parameters. In that setting, the thermodynamic metric and the covariance resolved by the detector become matrices, allowing one to examine how optimal feedback balances directions with different transport costs and degrees of observability. A complementary extension is to enlarge the control family, most simply by letting the stiffness respond to the observation as well, which would exploit a larger share of the information the measurement has already paid for~\cite{bauer2012efficiency}.

\acknowledgments{We thank Aidan Zentner and all other members of the Brenner Group for valuable feedback and discussions. The authors acknowledge the use of AI tools (Claude, ChatGPT, Gemini) to assist with analytical calculations, code and figure preparation, and manuscript editing. The authors are solely responsible for conceptualization and verification, and assume full responsibility for all contents of the present manuscript. This work was supported by NSF AI Institute of Dynamic Systems 2112085, the Harvard MRSEC (NSF DMR-2011754) and the Office of Naval Research through grant number ONR N00014-23-1-2654.}

\section*{Data Availability}
Analytical derivations are provided in the \textit{Supplementary Information}. Numerical scripts used to generate all figures are available at \url{https://github.com/fmottes/optimal-stepwise-feedback-control-paper}.

\bibliography{refs}


\clearpage
\onecolumngrid
\newgeometry{left=1.25in, right=1.25in, top=1in, bottom=1in}

\setcounter{equation}{0}
\setcounter{figure}{0}
\setcounter{table}{0}
\setcounter{section}{0}

\renewcommand{\theequation}{S\arabic{equation}}
\renewcommand{\thefigure}{S\arabic{figure}}
\renewcommand{\thetable}{S\arabic{table}}
\renewcommand{\thesection}{S\arabic{section}}
\renewcommand{\theHequation}{supp.\arabic{equation}}
\renewcommand{\theHfigure}{supp.\arabic{figure}}
\renewcommand{\theHtable}{supp.\arabic{table}}
\renewcommand{\theHsection}{supp.\arabic{section}}

\begin{center}
\textbf{\Large SUPPLEMENTARY INFORMATION}\\[0.7em]
\textbf{\large Optimal feedback control under stepwise equilibration and partial observation}\\[1em]
Francesco Mottes, Michael P. Brenner\\[0.5em]
\textit{School of Engineering and Applied Sciences, Harvard University, Cambridge MA 02138}
\end{center}
\vspace{1em}

\section{Bellman recursion for minimum expected work}
\label{sec:supp-theory}

We derive the Bellman recursion underlying the main text directly in the partial-observation setting. The fully observed case is recovered as the special case $p(o\mid x,H)=\delta(o-x)$, which collapses the posterior to a Dirac mass at the true microstate; no separate derivation is needed.

\subsection{Protocol and factorized path measure}

The protocol consists of $N$ feedback cycles indexed by $n=0,1,\dots,N-1$. Each cycle comprises three sub-steps: (i) equilibration of the system under the current Hamiltonian $H_n$, (ii) acquisition of an observation $o_n$ from a measurement kernel $p(o_n\mid x_n,H_n)$, and (iii) an instantaneous quench $H_n\to H_{n+1}$ selected by the controller. The initial Hamiltonian $H_0$ is prescribed and the protocol is constrained to terminate at a prescribed target $H_N=H_f$.

Equilibration under a fixed Hamiltonian exchanges heat with the bath but performs no work, so all protocol work is done during the instantaneous quenches. Equilibration also erases history: the post-equilibration microstate $x_n$ depends on the past only through $H_n$. Consequently the pair $(H_n,o_n)$ is a sufficient statistic for decision-making, and optimal policies may be taken to depend only on $(o_n,H_n)$ rather than on the full history $(H_0,o_0,\dots,H_n,o_n)$.

After equilibration, the microstate is drawn from the Boltzmann distribution
\begin{equation}
p_H^{\mathrm{eq}}(x)=\frac{e^{-\beta H(x)}}{Z[H]},
\qquad
Z[H]=\int dx\, e^{-\beta H(x)},
\label{eq:supp-boltzmann}
\end{equation}
with $\beta=(k_\mathrm{B}T)^{-1}$. The controller then draws $o_n\sim p(\cdot\mid x_n,H_n)$ and selects the next Hamiltonian from a feedback policy $\pi(H_{n+1}\mid o_n,H_n)$. The one-step factorized path measure is
\begin{equation}
 p(x_n,o_n,H_{n+1}\mid H_n)
 =
 p_{H_n}^{\mathrm{eq}}(x_n)\,p(o_n\mid x_n,H_n)\,\pi(H_{n+1}\mid o_n,H_n),
\label{eq:supp-path-measure}
\end{equation}
and for a deterministic policy this reduces to $H_{n+1}=\phi(o_n;H_n)$. Because each quench is instantaneous, the microstate is held fixed during the intervention and the stochastic work increment is
\begin{equation}
 W_n = H_{n+1}(x_n)-H_n(x_n).
\label{eq:supp-work-step}
\end{equation}
The total work along the trajectory is additive, $W(\tau)=\sum_{n=0}^{N-1}[H_{n+1}(x_n)-H_n(x_n)]$, where successive $x_n$ are \emph{independent} samples drawn from the corresponding equilibrium distributions $p_{H_n}^{\mathrm{eq}}$ (re-equilibration between quenches decorrelates the microstate). These two facts---factorized path measure and additive work---are the only ingredients needed below.

\subsection{Posterior distribution and Bellman recursion}

We write all averages in continuous-state form, with $\int dx$, $\int do$, and $\int dH'$ over the microstate, observation, and Hamiltonian spaces; each reduces to a sum when the corresponding space is discrete, with $\pi(\cdot\mid o,H)$ a probability mass function over actions. Given the current Hamiltonian $H$ and an observation $o$, Bayes' rule gives the posterior distribution of the microstate,
\begin{equation}
 p(x\mid o,H)
 =
 \frac{p(o\mid x,H)\,p_H^{\mathrm{eq}}(x)}{\int dx'\,p(o\mid x',H)\,p_H^{\mathrm{eq}}(x')},
\label{eq:supp-posterior}
\end{equation}
with marginal observation law
\begin{equation}
 p(o\mid H)=\int dx\,p(o\mid x,H)\,p_H^{\mathrm{eq}}(x).
\label{eq:supp-observation-marginal}
\end{equation}
Let $V_r(H)$ denote the minimum expected remaining work needed to reach the target Hamiltonian $H_f$ in exactly $r$ additional quenches, starting from equilibrium under $H$. The terminal condition is
\begin{equation}
V_0(H)=
\begin{cases}
0, & H=H_f,\\
+\infty, & H\neq H_f.
\end{cases}
\label{eq:supp-terminal}
\end{equation}
The conditional expected immediate work of choosing $H'$ after observing $o$ is
\begin{equation}
w(H,o;H'):=\mathbb E_{x\sim p(x\mid o,H)}\bigl[H'(x)-H(x)\bigr],
\label{eq:supp-stagecost}
\end{equation}
We first evaluate the cost of an arbitrary---not necessarily optimal---policy $\pi$. Following $\pi$ for one more cycle from equilibrium under $H$ incurs the immediate work $w(H,o;H')$ of the quench $H\to H'$ and then leaves the controller facing the same problem with one fewer quench remaining. 
Assume the minimum remaining cost from any $H'$ is already known and equals $V_{r-1}(H')$. Because re-equilibration under $H'$ erases the microstate history, this does not depend on what action is chosen at step $r$.
The expected $r$-step cost of $\pi$ is this immediate-plus-continuation sum, averaged over the two sources of randomness the controller faces this cycle: the observation $o$, drawn from its marginal $p(o\mid H)$, and, given $o$, the action $H'$, drawn from the policy $\pi(\cdot\mid o,H)$,
\begin{equation}
J_r^{\pi}(H)
=
\int do\,p(o\mid H)
\int dH'\,\pi(H'\mid o,H)
\bigl[w(H,o;H')+V_{r-1}(H')\bigr].
\label{eq:supp-Jn}
\end{equation}

The value function is the least cost achievable over all policies, $V_r(H)=\inf_{\pi}J_r^{\pi}(H)$. This looks like a minimization over an entire function $\pi(\cdot\mid\cdot,H)$, but it collapses to a pointwise one: the action chosen in response to one observation places no constraint on the action chosen in response to another---distinct observations are decoupled---and the weight $p(o\mid H)$ is nonnegative, so the observation integral is minimized by minimizing its integrand separately at each $o$. The infimum therefore moves inside the observation average,
\begin{equation}
V_r(H)
=
\int do\,p(o\mid H)
\inf_{\pi(\cdot\mid o,H)}
\int dH'\,\pi(H'\mid o,H)
\bigl[w(H,o;H')+V_{r-1}(H')\bigr].
\label{eq:supp-bellman-pre}
\end{equation}

\subsection{Optimality of deterministic policies}

For fixed $(H,o)$ define
\begin{equation}
Q_r(H,o;H'):=w(H,o;H')+V_{r-1}(H').
\label{eq:supp-Q}
\end{equation}
The inner objective $\int dH'\,\pi(H'\mid o,H)\,Q_r(H,o;H')$ in \eqref{eq:supp-bellman-pre} is linear in the policy $\pi(\cdot\mid o,H)$, so its infimum is attained at a point mass on a minimizing action: an average cannot be lower than its smallest term, and randomization cannot improve on the best deterministic choice~\cite{bertsekas2012dynamic}. Hence an optimal policy may be taken deterministic,
\begin{equation}
\pi^*(H'\mid o,H)=\delta\bigl(H'-H_r^*(H,o)\bigr),
\qquad
H_r^*(H,o)\in \arg\min_{H'} Q_r(H,o;H'),
\label{eq:supp-deterministic-policy}
\end{equation}
with $\delta$ the Dirac mass (a Kronecker delta for a discrete action family); for a continuous family the minimizer is assumed to exist, as it does for the harmonic family of Section~\ref{sec:supp-pomdp}. Substituting \eqref{eq:supp-deterministic-policy} into \eqref{eq:supp-bellman-pre} gives the exact Bellman recursion
\begin{equation}
\boxed{
V_r(H)
=
\mathbb E_{o\sim p(o\mid H)}
\left[
\min_{H'}\bigl(w(H,o;H')+V_{r-1}(H')\bigr)
\right].
}
\label{eq:supp-bellman}
\end{equation}
Full observation is recovered by setting $p(o\mid x,H)=\delta(o-x)$: then $p(x\mid o,H)=\delta(x-o)$ and $w(H,o;H')=H'(o)-H(o)$, so Eq.~\eqref{eq:supp-bellman} becomes
\begin{equation}
V_r(H)=\mathbb E_{x\sim p_H^{\mathrm{eq}}}\!\left[\min_{H'}\bigl(H'(x)-H(x)+V_{r-1}(H')\bigr)\right].
\label{eq:supp-bellman-fullobs}
\end{equation}

\section{Harmonic translated-trap family under partial observation}
\label{sec:supp-pomdp}

We now specialize to the translated harmonic family
\begin{equation}
H_{\lambda}(x)=\frac{k}{2}(x-\lambda)^2,
\qquad k>0,
\label{eq:supp-harmonic}
\end{equation}
with target $\lambda=L$. The equilibrium density at trap center $\lambda$ is
\begin{equation}
p_\lambda^{\mathrm{eq}}(x)=\frac{\exp[-\beta k(x-\lambda)^2/2]}{Z_\lambda}.
\label{eq:supp-plambda}
\end{equation}
The change of variables $y=x-\lambda$ gives
\begin{equation}
Z_\lambda=\int_{-\infty}^{\infty}e^{-\beta k y^2/2}\,dy=\sqrt{\frac{2\pi}{\beta k}},
\label{eq:supp-partition}
\end{equation}
independent of $\lambda$. The equilibrium free energy $F_\lambda=-\beta^{-1}\ln Z_\lambda$ is therefore $\lambda$-independent, and the equilibrium distribution is Gaussian:
\begin{equation}
x\sim \mathcal N(\lambda,\sigma_{\mathrm{eq}}^2),
\qquad
\sigma_{\mathrm{eq}}^2=\frac{1}{\beta k}=\frac{k_{\mathrm B}T}{k}.
\label{eq:supp-eq-gaussian}
\end{equation}
The terminal Hamiltonian is $H_f=H_L$, so the terminal condition of Section~\ref{sec:supp-theory} specializes to
\begin{equation}
V_0(\lambda)=\begin{cases}0,& \lambda=L,\\ +\infty,& \lambda\neq L.\end{cases}
\label{eq:supp-terminal-harmonic}
\end{equation}
Because the family is labelled by the trap center alone, we write $\lambda$ for $H_\lambda$ in function arguments, so that $V_r(\lambda)\equiv V_r(H_\lambda)$. Let $V_r(\lambda)$ denote the minimal expected remaining work to reach $L$ in exactly $r$ additional quenches, starting from equilibrium at trap center $\lambda$; the overall $N$-cycle protocol starts at $\lambda_0=0$ and the quantity of interest is $V_N(0)$.

\subsection{Translation-covariant observation kernels}
\label{sec:supp-translation-covariance}

The closed form below rests on a single assumption: the detector shares the translation symmetry of the harmonic family. Shifting the trap, the particle, and the readout together by a common amount $a$ must leave the measurement statistics unchanged,
\begin{equation}
p(T_a o\mid x+a,\,\lambda+a)=p(o\mid x,\lambda),
\label{eq:supp-tc-def}
\end{equation}
where $T_a$ is the induced shift on the observation---an actual translation $T_a o=o+a$ for a detector that reports a spatial coordinate, or the identity for one whose output is already relative to the trap center. This covers essentially every realistic read-out: additive-noise position detectors $o=x+\xi$ with any fixed noise law (Gaussian, Laplace, Cauchy, and heavier-tailed variants), and trap-relative discrete detectors whose statistics depend only on $x-\lambda$ (sign, threshold, trap-centered bins, photon counts). It fails only for detectors keyed to an absolute lab location, such as absolute-position bins or thresholds, or multiplicative noise $o=\alpha x$.

The point of the assumption is that it makes the inference problem the same at every trap position. Viewed in trap-centered coordinates $u=x-\lambda$, the prior of $u$ is the fixed Gaussian $\mathcal N(0,\sigma_{\mathrm{eq}}^2)$ and, by covariance, the law of the trap-relative observation is fixed too; the posterior---and hence the estimator built from it---then depends on $\lambda$ only through a rigid offset. The microstate estimate is therefore unbiased and tracks the trap, and its spread across observations is a single $\lambda$-independent constant,
\begin{equation}
\hat{x}(o;\lambda):=\mathbb E[x\mid o,\lambda],
\qquad
\mathbb E[\hat{x}\mid\lambda]=\lambda,
\qquad
\eta^2:=\operatorname{Var}(\hat{x}\mid\lambda)\ \ \text{(independent of $\lambda$)}.
\label{eq:supp-tc-moments}
\end{equation}
The law of total variance, $\sigma_{\mathrm{eq}}^2=\mathbb E[\operatorname{Var}(x\mid o,\lambda)\mid\lambda]+\eta^2$, bounds $\eta^2\in[0,\sigma_{\mathrm{eq}}^2]$, with the upper end reached under full observation ($\hat{x}=x$) and the lower end by a useless channel ($\hat{x}\equiv\lambda$). The channel's usefulness thus collapses to a single dial, the resolved-variance fraction $\chi:=\eta^2/\sigma_{\mathrm{eq}}^2\in[0,1]$. This $\lambda$-independence is precisely what the closed form needs: for channels that violate Eq.~\eqref{eq:supp-tc-def}, $\eta^2$ becomes $\lambda$-dependent, the $a_r$ and $b_r$ recursions derived below no longer decouple, and the solution does not apply.

\subsection{Posterior-averaged stage cost}

Fix the current center $\lambda$, an observation $o$, and a candidate next center $\lambda'$. The conditional expected immediate work is
\begin{equation}
w(\lambda,o;\lambda'):=\mathbb E[W_n\mid o,\lambda,\lambda']
=
\mathbb E\!\left[\tfrac{k}{2}(x-\lambda')^2-\tfrac{k}{2}(x-\lambda)^2\,\middle|\,o,\lambda\right].
\label{eq:supp-stage-harmonic-start}
\end{equation}
The cancellation can be seen directly at the level of the Hamiltonian difference. Because the quench translates the trap without changing its stiffness,
\begin{align}
H_{\lambda'}(x)-H_\lambda(x)
&=
\frac{k}{2}(x-\lambda')^2-\frac{k}{2}(x-\lambda)^2
\notag\\
&=
k(\lambda-\lambda')x+\frac{k}{2}(\lambda'^2-\lambda^2).
\label{eq:supp-harmonic-difference}
\end{align}
The $x^2$ term cancels pointwise, leaving a difference \emph{linear} in $x$. Taking the conditional expectation of this expression replaces $x$ by the posterior estimate $\hat{x}(o;\lambda):=\mathbb E[x\mid o,\lambda]$. Re-completing the square in $\hat{x}$ gives
\begin{equation}
w(\lambda,o;\lambda')
=
\frac{k}{2}(\hat{x}-\lambda')^2-\frac{k}{2}(\hat{x}-\lambda)^2.
\label{eq:supp-stage-harmonic}
\end{equation}
The Bellman equation then becomes
\begin{equation}
V_r(\lambda)
=
\mathbb E_{o\sim p(o\mid \lambda)}
\!\left[
\min_{\lambda'}
\left(
\tfrac{k}{2}(\hat{x}(o;\lambda)-\lambda')^2
-\tfrac{k}{2}(\hat{x}(o;\lambda)-\lambda)^2
+V_{r-1}(\lambda')
\right)
\right],
\label{eq:supp-bellman-harmonic}
\end{equation}
This has exactly the same form as the fully observed harmonic Bellman equation, with the measured microstate $x$ replaced by its estimate $\hat{x}$.

\subsection{Value function and minimum work}

We seek a value function of the form
\begin{equation}
V_r(\lambda)=a_r(\lambda-L)^2+b_r.
\label{eq:supp-ansatz}
\end{equation}

\textbf{Recursion for the coefficients.} The term $-k(\hat{x}-\lambda)^2/2$ in Eq.~\eqref{eq:supp-bellman-harmonic} is independent of $\lambda'$, so we may first minimize the remaining terms pointwise for each $\hat{x}$ and then substitute their minimum back into the outer expectation to obtain $V_r$.

Assume inductively that $V_{r-1}(\lambda)=a_{r-1}(\lambda-L)^2+b_{r-1}$. For a fixed estimate $\hat{x}$, the part of the Bellman objective that depends on the next trap center $\lambda'$ is
\begin{equation}
g(\lambda';\hat{x})=\tfrac{k}{2}(\hat{x}-\lambda')^2+V_{r-1}(\lambda')=\tfrac{k}{2}(\hat{x}-\lambda')^2+a_{r-1}(\lambda'-L)^2+b_{r-1}.
\label{eq:supp-g}
\end{equation}
Differentiating with respect to $\lambda'$ gives
\begin{equation}
\frac{\partial g}{\partial\lambda'}
=k(\lambda'-\hat{x})+2a_{r-1}(\lambda'-L).
\end{equation}
Since $\partial^2g/\partial\lambda'^2=k+2a_{r-1}>0$, setting the first derivative to zero gives the unique minimizer
\begin{equation}
(k+2a_{r-1})\lambda'=k\hat{x}+2a_{r-1}L.
\label{eq:supp-stationary}
\end{equation}
Thus the optimal deterministic action is
\begin{equation}
\lambda^*(o;r)=\frac{k\,\hat{x}(o;\lambda)+2a_{r-1}L}{k+2a_{r-1}}.
\label{eq:supp-policy-general}
\end{equation}
At this minimizer, the two displacements in \eqref{eq:supp-g} are
\begin{equation}
\hat{x}-\lambda^*=\frac{2a_{r-1}}{k+2a_{r-1}}(\hat{x}-L),
\qquad
\lambda^*-L=\frac{k}{k+2a_{r-1}}(\hat{x}-L).
\label{eq:supp-optimal-displacements}
\end{equation}
Substitution into \eqref{eq:supp-g} therefore gives
\begin{align}
g_{\min}(\hat{x})
&=
\left[
\frac{k}{2}\left(\frac{2a_{r-1}}{k+2a_{r-1}}\right)^2
+a_{r-1}\left(\frac{k}{k+2a_{r-1}}\right)^2
\right](\hat{x}-L)^2+b_{r-1}
\notag\\
&=\frac{k a_{r-1}}{k+2a_{r-1}}(\hat{x}-L)^2+b_{r-1}.
\label{eq:supp-gmin}
\end{align}
Substituting into \eqref{eq:supp-bellman-harmonic} gives
\begin{align}
V_r(\lambda)
&=
\mathbb E_{p(o\mid\lambda)}\!\left[
g_{\min}(\hat{x})-\frac{k}{2}(\hat{x}-\lambda)^2
\right].
\label{eq:supp-Vr-expectation}
\end{align}
Recall that $\eta^2:=\operatorname{Var}(\hat{x}\mid\lambda) = \mathbb E_{p(o\mid\lambda)}[(\hat{x}-\lambda)^2]$. Using
\begin{equation}
\mathbb E_{p(o\mid\lambda)}[(\hat{x}-L)^2]
=(\lambda-L)^2+\eta^2,
\label{eq:supp-estimate-moments}
\end{equation}
we obtain
\begin{align}
V_r(\lambda)
&=
\frac{k a_{r-1}}{k+2a_{r-1}}(\lambda-L)^2+b_{r-1}
-\left(
\frac{k}{2}-\frac{k a_{r-1}}{k+2a_{r-1}}
\right)\eta^2
\notag\\
&=
\frac{k a_{r-1}}{k+2a_{r-1}}(\lambda-L)^2+b_{r-1}
-\frac{k^2\eta^2}{2(k+2a_{r-1})}.
\label{eq:supp-Vr-recursion}
\end{align}
Matching this expression with \eqref{eq:supp-ansatz} identifies the recursions
\begin{equation}
a_r=\frac{k\,a_{r-1}}{k+2a_{r-1}},
\label{eq:supp-a-recursion}
\end{equation}
and
\begin{equation}
b_r=b_{r-1}-\frac{k^2\eta^2}{2(k+2a_{r-1})}.
\label{eq:supp-b-recursion}
\end{equation}

\textbf{Base case ($r=1$).} We need to combine this recursion with the base case, which is $r=1$. Because $V_0(\lambda')=+\infty$ for $\lambda'\neq L$ (Eq.~\eqref{eq:supp-terminal-harmonic}), the inner minimum in the Bellman equation~\eqref{eq:supp-bellman-harmonic} at $r=1$ is finite only when $\lambda'=L$; the final quench is therefore deterministically forced to $\lambda'=L$, and
\begin{equation}
V_1(\lambda)
=
\mathbb E_{o\sim p(o\mid \lambda)}
\!\left[
\tfrac{k}{2}(\hat{x}(o;\lambda)-L)^2-\tfrac{k}{2}(\hat{x}(o;\lambda)-\lambda)^2
\right].
\label{eq:supp-V1-start}
\end{equation}
Using $\mathbb E[\hat{x}\mid \lambda]=\lambda$, we obtain
\begin{equation}
V_1(\lambda)=\frac{k}{2}(\lambda-L)^2,
\qquad
a_1=\frac{k}{2},\qquad b_1=0.
\label{eq:supp-V1}
\end{equation}

\textbf{Closed forms for the coefficients.} We first solve the recursion for $a_r$. Introduce the dimensionless coefficient $c_r$ by
\begin{equation}
a_r=\frac{k}{2c_r}.
\label{eq:supp-c-definition}
\end{equation}
Substituting $a_{r-1}=k/(2c_{r-1})$ into \eqref{eq:supp-a-recursion} gives
\begin{align}
a_r
&=\frac{k\,[k/(2c_{r-1})]}
{k+2[k/(2c_{r-1})]}
=\frac{k^2/(2c_{r-1})}
{k(c_{r-1}+1)/c_{r-1}}
=\frac{k}{2(c_{r-1}+1)}.
\label{eq:supp-a-c-substitution}
\end{align}
Comparing this result with $a_r=k/(2c_r)$ shows that
\begin{equation}
c_r=c_{r-1}+1.
\end{equation}
The base value $a_1=k/2$ corresponds to $c_1=1$. Iterating the recursion therefore gives
\begin{equation}
c_r
=c_1+\sum_{q=2}^{r}(c_q-c_{q-1})
=1+\sum_{q=2}^{r}1
=1+(r-1)
=r,
\end{equation}
and hence
\begin{equation}
a_r=\frac{k}{2r}.
\label{eq:supp-a-closed}
\end{equation}

We next substitute $a_{r-1}=k/[2(r-1)]$ into the recursion for $b_r$. For $r\geq2$,
\begin{align}
b_r-b_{r-1}
&=-\frac{k^2\eta^2}{2(k+2a_{r-1})}
=-\frac{k^2\eta^2}
{2\left(k+k/(r-1)\right)}
=-\frac{k\eta^2}{2}\,\frac{r-1}{r}.
\label{eq:supp-b-step}
\end{align}
The increments telescope from the base value $b_1=0$:
\begin{align}
b_r
&=b_1+\sum_{q=2}^{r}(b_q-b_{q-1})
=-\frac{k\eta^2}{2}\sum_{q=2}^{r}\frac{q-1}{q}
=-\frac{k\eta^2}{2}
\left[(r-1)-\sum_{q=2}^{r}\frac{1}{q}\right].
\label{eq:supp-b-sum}
\end{align}
Defining the harmonic number $\mathcal{H}_r:=\sum_{q=1}^{r}1/q$, we have
\begin{equation}
(r-1)-\sum_{q=2}^{r}\frac{1}{q}
=(r-1)-(\mathcal{H}_r-1)
=r-\mathcal{H}_r.
\end{equation}
Therefore
\begin{equation}
b_r=-\frac{k\eta^2}{2}\bigl(r-\mathcal{H}_r\bigr).
\label{eq:supp-b-closed}
\end{equation}

\textbf{Value function.} The value function is therefore
\begin{equation}
V_r(\lambda)=\frac{k}{2r}(\lambda-L)^2-\frac{k\eta^2}{2}\bigl(r-\mathcal{H}_r\bigr).
\label{eq:supp-value-final}
\end{equation}
Starting from $\lambda_0=0$ with $N$ steps remaining, the minimum expected work is
\begin{equation}
\boxed{
W_N^*=V_N(0)=\frac{kL^2}{2N}-\frac{k\eta^2}{2}\bigl(N-\mathcal{H}_N\bigr).
}
\label{eq:supp-min-work}
\end{equation}
Using the resolved-variance fraction $\chi=\eta^2/\sigma_{\mathrm{eq}}^2\in[0,1]$ and $\sigma_{\mathrm{eq}}^2=k_{\mathrm B}T/k$, this also reads
\begin{equation}
\boxed{
W_N^*=\frac{kL^2}{2N}-\frac{k_{\mathrm B}T}{2}\,\chi\,\bigl(N-\mathcal{H}_N\bigr).
}
\label{eq:supp-min-work-chi}
\end{equation}

\subsection{Optimal feedback law}

The optimal action at each observation is the minimizer $\lambda^*(o;r)$ of the stage-plus-continuation cost $g$, derived in \eqref{eq:supp-policy-general}. Substituting the now-solved coefficient $a_{r-1}=k/[2(r-1)]$ into that expression yields the explicit feedback law
\begin{equation}
\boxed{
\lambda^*(o;r)=\frac{r-1}{r}\,\hat{x}(o;\lambda)+\frac{1}{r}\,L.
}
\label{eq:supp-policy-main}
\end{equation}
For long remaining horizons ($r$ large) the controller tracks the estimate, whereas for $r=1$ it is forced onto the terminal target $L$, interpolating smoothly between the two regimes.

\subsection{Fully observed limit}
The fully observed case corresponds to $p(o\mid x,H_\lambda)=\delta(o-x)$, for which the posterior is a Dirac mass at $x$, so $\hat{x}(o;\lambda)=x$, $\eta^2=\sigma_{\mathrm{eq}}^2$, and $\chi=1$. Equations~\eqref{eq:supp-policy-main} and \eqref{eq:supp-min-work-chi} then reduce to
\begin{equation}
\lambda^*(x;r)=\frac{r-1}{r}\,x+\frac{1}{r}L,
\qquad
W_N^*\big|_{\chi=1}=\frac{kL^2}{2N}-\frac{k_{\mathrm B}T}{2}\bigl(N-\mathcal{H}_N\bigr),
\label{eq:supp-full-obs-limit}
\end{equation}
so the partial-observation solution contains the fully observed closed form as a special case.

\section{Gaussian observation channel}
\label{sec:supp-gaussian}

Consider an observation channel with additive Gaussian measurement noise,
\begin{equation}
o=x+\xi,
\qquad
\xi\sim \mathcal N(0,\Delta^2),
\label{eq:supp-gaussian-noise}
\end{equation}
with $\xi$ independent of $x$. Under equilibration at the current center, $x\mid\lambda\sim\mathcal N(\lambda,\sigma_{\mathrm{eq}}^2)$, and hence
\begin{equation}
o\mid\lambda\sim
\mathcal N\!\left(\lambda,\sigma_{\mathrm{eq}}^2+\Delta^2\right).
\label{eq:supp-gaussian-marginal}
\end{equation}
Bayes' rule gives the posterior density, up to normalization, as
\begin{align}
p(x\mid o,\lambda)
&\propto p(o\mid x)\,p(x\mid\lambda)
\propto
\exp\!\left[
-\frac{(o-x)^2}{2\Delta^2}
-\frac{(x-\lambda)^2}{2\sigma_{\mathrm{eq}}^2}
\right].
\label{eq:supp-gaussian-posterior-density}
\end{align}
Collecting the terms that depend on $x$, the exponent becomes
\begin{equation}
-\frac{1}{2}
\left[
\left(\frac{1}{\sigma_{\mathrm{eq}}^2}+\frac{1}{\Delta^2}\right)x^2
-2\left(
\frac{\lambda}{\sigma_{\mathrm{eq}}^2}+\frac{o}{\Delta^2}
\right)x
\right]
+\text{const}.
\label{eq:supp-gaussian-posterior-exponent}
\end{equation}
This is the exponent of a Gaussian whose mean is
\begin{align}
\hat{x}(o;\lambda)
&=\frac{\lambda/\sigma_{\mathrm{eq}}^2+o/\Delta^2}
{1/\sigma_{\mathrm{eq}}^2+1/\Delta^2}
=\frac{\Delta^2\lambda+\sigma_{\mathrm{eq}}^2o}
{\sigma_{\mathrm{eq}}^2+\Delta^2}
=\lambda+\rho\,(o-\lambda),
\qquad
\rho:=\frac{\sigma_{\mathrm{eq}}^2}
{\sigma_{\mathrm{eq}}^2+\Delta^2}.
\label{eq:supp-gaussian-posterior-mean}
\end{align}
Substituting the posterior mean \eqref{eq:supp-gaussian-posterior-mean} into the general feedback law \eqref{eq:supp-policy-main} gives
\begin{equation}
\boxed{
\lambda^*(o;r)
=\frac{r-1}{r}\bigl[\lambda+\rho\,(o-\lambda)\bigr]+\frac{1}{r}L,
}
\label{eq:supp-gaussian-policy}
\end{equation}
Because the estimator is affine in $o$, its variance across observations is
\begin{equation}
\eta^2
=\operatorname{Var}\!\bigl(\hat{x}(o;\lambda)\mid \lambda\bigr)
=\rho^2\operatorname{Var}(o\mid\lambda)
=\rho^2(\sigma_{\mathrm{eq}}^2+\Delta^2)
=\rho\,\sigma_{\mathrm{eq}}^2,
\label{eq:supp-eta-gaussian}
\end{equation}
where the last equality follows from the definition of $\rho$. Thus the resolved-variance fraction is simply
\begin{equation}
\chi=\frac{\eta^2}{\sigma_{\mathrm{eq}}^2}=\rho.
\label{eq:supp-chi-gaussian}
\end{equation}
Substituting $\chi=\rho$ into \eqref{eq:supp-min-work-chi} similarly gives the minimum expected work
\begin{equation}
\boxed{
W_{N,\mathrm{Gauss}}^*=\frac{kL^2}{2N}-\frac{k_{\mathrm B}T}{2}\,\rho\,\bigl(N-\mathcal{H}_N\bigr).
}
\label{eq:supp-gaussian-work}
\end{equation}
These expressions satisfy the expected limits:
\begin{itemize}
\item as $\Delta^2\to 0$, $\rho\to 1$ and the full-information formulas \eqref{eq:supp-full-obs-limit} are recovered;
\item as $\Delta^2\to \infty$, $\rho\to 0$, the feedback gain disappears, and one is left with the open-loop transport cost $W_N^*=kL^2/(2N)$ of the next section.
\end{itemize}

\subsection{Information cost and saturation fraction}

The work results above depend on the Gaussian channel only through the resolved-variance fraction $\chi=\rho$. Its information-theoretic cost, by contrast, is measured by the mutual information $I(x;o)=H(o)-H(o\mid x)$. Equation~\eqref{eq:supp-gaussian-marginal} gives $H(o)=\tfrac{1}{2}\ln[2\pi e(\sigma_{\mathrm{eq}}^2+\Delta^2)]$. Conditioning on $x$ leaves only the measurement noise, so $H(o\mid x)=\tfrac{1}{2}\ln(2\pi e\,\Delta^2)$. Their difference is
\begin{equation}\label{eq:mutual-information}
I(x;o)=\tfrac{1}{2}\ln\frac{\sigma_{\mathrm{eq}}^2+\Delta^2}{\Delta^2}=\tfrac{1}{2}\ln\frac{1}{1-\chi},
\end{equation}
where the second equality uses $\chi=\rho$ together with the definition of $\rho$ in Eq.~\eqref{eq:supp-gaussian-posterior-mean}. This is a function of $\chi$ alone that diverges as $\chi\to1$, when the observation pins down the microstate.

\begin{figure}[t]
\centering
\includegraphics[width=.8\columnwidth]{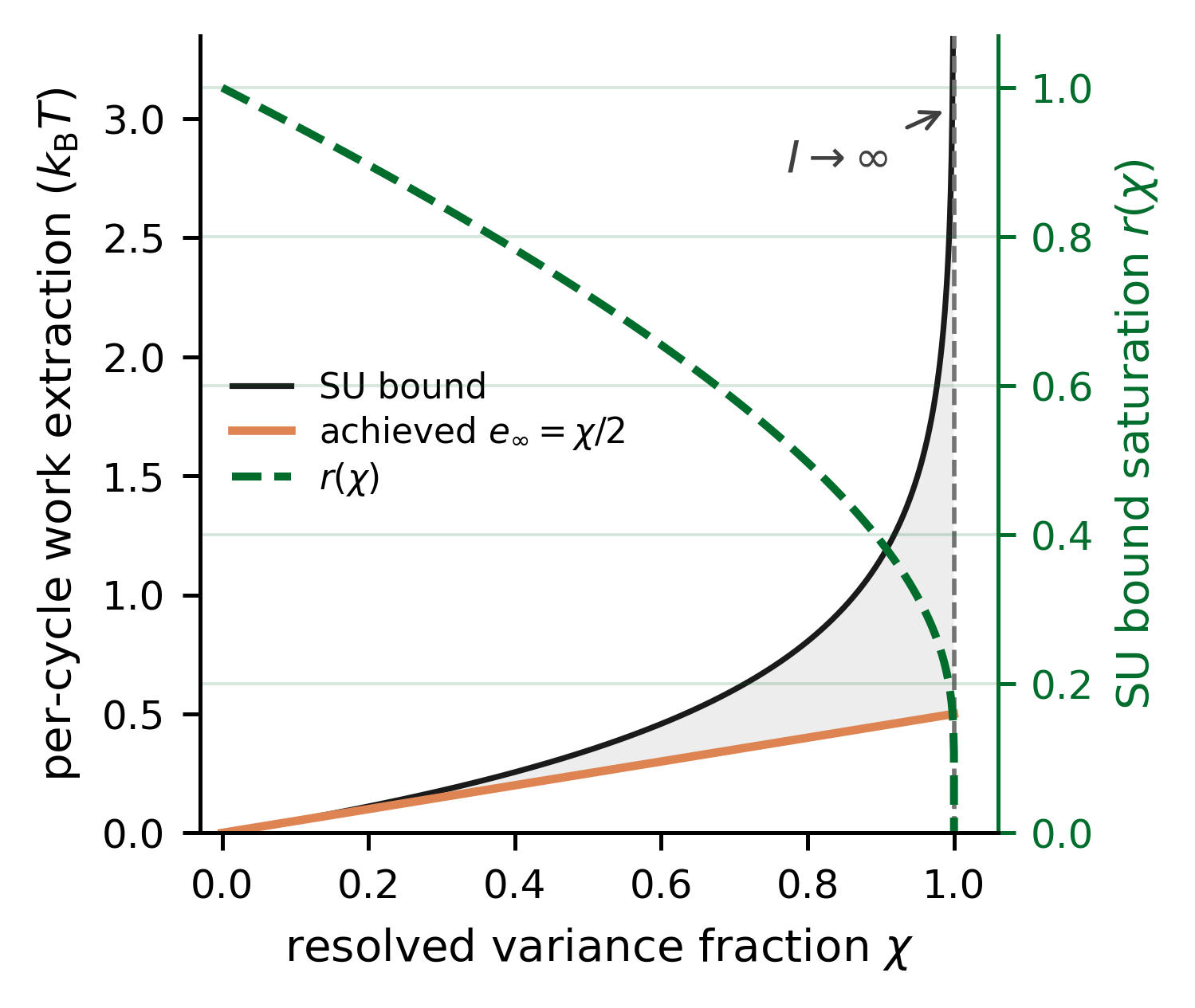}
\caption{Information-thermodynamic consistency for the Gaussian observation channel. The left axis compares the per-cycle Sagawa--Ueda bound $I(x;o)=\tfrac{1}{2}\ln[1/(1-\chi)]$ with the asymptotic extraction achieved by translated harmonic quenches, $e_\infty=\chi/2$, both in units of $k_{\mathrm B}T$. The shaded region is the unused information gap. The right axis shows the saturation fraction $r(\chi)=\chi/\ln[1/(1-\chi)]$, which approaches one for weak measurements and vanishes as the bound diverges near perfect observation.}
\label{fig:information-consistency}
\end{figure}

The optimal protocol's asymptotic extraction is $-W_N^*/N\to\chi k_{\mathrm B}T/2$ per cycle as $N\to\infty$. Dividing by the information bound $k_{\mathrm B}T I(x;o)$ gives the saturation fraction
\begin{equation}\label{eq:saturation-ratio}
r(\chi)
:=\frac{\chi/2}{I(x;o)}
=\frac{\chi}{\ln[1/(1-\chi)]}
\in(0,1],
\end{equation}
falling from $r(\chi)\to1$ at $\chi\to0$ to $r(\chi)\to0$ at $\chi\to1$. Purely translational harmonic quenches therefore saturate the bound at weak measurement but leave a growing share of the information unused as the measurement sharpens (Fig.~\ref{fig:information-consistency}). The gap arises because the controller can only shift the trap center without reshaping the distribution, and therefore under-utilizes available information at high state resolution.

\textbf{General measurement channel.} Although Eq.~\eqref{eq:mutual-information} was derived for the Gaussian channel, it bounds every channel with the same resolved fraction. Fixing $\chi$ fixes the mean-square error $\mathbb E[(x-\hat{x})^2]=(1-\chi)\sigma_{\mathrm{eq}}^2$ of the estimate $\hat{x}=\hat{x}(o;\lambda)$. Because $\hat{x}$ is a function of $o$, the data-processing inequality and the Gaussian rate--distortion theorem~\cite{cover1999elements} give
\begin{equation}\label{eq:rate-distortion-bound}
I(x;o)\ge I(x;\hat{x})
\ge\tfrac{1}{2}\ln\frac{\sigma_{\mathrm{eq}}^2}{(1-\chi)\sigma_{\mathrm{eq}}^2}
=\tfrac{1}{2}\ln\frac{1}{1-\chi},
\end{equation}
with equality for the additive Gaussian channel. The Gaussian channel therefore minimizes the mutual information at fixed $\chi$, so Eq.~\eqref{eq:mutual-information} is the least possible Landauer floor---and, correspondingly, $r(\chi)$ is the largest saturation fraction---over all channels of that resolved fraction.

\section{Open-loop comparison}
\label{sec:supp-open-loop}

For comparison, consider a feedback-free protocol with prescribed trap centers
$\lambda_0=0,\lambda_1,\ldots,\lambda_N=L$. Define the displacement of the
$i$th quench by
\begin{equation}
\Delta\lambda_i:=\lambda_i-\lambda_{i-1},
\qquad
\sum_{i=1}^{N}\Delta\lambda_i=L.
\label{eq:supp-open-loop-constraint}
\end{equation}
Before this quench, equilibration at $\lambda_{i-1}$ gives
$x_i\sim\mathcal N(\lambda_{i-1},\sigma_{\mathrm{eq}}^2)$, and therefore
$\mathbb E[x_i-\lambda_{i-1}]=0$.
The expected work performed by shifting the center from $\lambda_{i-1}$ to
$\lambda_i=\lambda_{i-1}+\Delta\lambda_i$ is therefore
\begin{align}
\mathbb E[W_i]
&=\frac{k}{2}\,
\mathbb E\!\left[
(x_i-\lambda_i)^2-(x_i-\lambda_{i-1})^2
\right]
\notag\\
&=\frac{k}{2}\,
\mathbb E\!\left[
(x_i-\lambda_{i-1}-\Delta\lambda_i)^2
-(x_i-\lambda_{i-1})^2
\right]
\notag\\
&=\frac{k}{2}\left[
-2\Delta\lambda_i\,\mathbb E[x_i-\lambda_{i-1}]
+(\Delta\lambda_i)^2
\right]
\notag\\
&=\frac{k}{2}(\Delta\lambda_i)^2.
\label{eq:supp-open-loop-start}
\end{align}
Consequently, the total expected work of an arbitrary open-loop schedule is
\begin{equation}
W_N^{\mathrm{open}}(\Delta\lambda_1,\ldots,\Delta\lambda_N)
=\frac{k}{2}\sum_{i=1}^{N}(\Delta\lambda_i)^2.
\label{eq:supp-open-loop-arbitrary}
\end{equation}
To minimize this expression under the displacement constraint, subtract the
mean increment $L/N$ from each $\Delta\lambda_i$. Expanding the resulting
sum of squares gives
\begin{align}
\sum_{i=1}^{N}\left(\Delta\lambda_i-\frac{L}{N}\right)^2
&=\sum_{i=1}^{N}(\Delta\lambda_i)^2
-\frac{2L}{N}\sum_{i=1}^{N}\Delta\lambda_i
+N\left(\frac{L}{N}\right)^2
=\sum_{i=1}^{N}(\Delta\lambda_i)^2-\frac{L^2}{N},
\label{eq:supp-open-loop-decomposition}
\end{align}
where we used $\sum_i\Delta\lambda_i=L$. The left-hand side is
nonnegative, and therefore
\begin{equation}
\sum_{i=1}^{N}(\Delta\lambda_i)^2\geq\frac{L^2}{N}.
\end{equation}
Equality holds if and only if every squared deviation vanishes, or equivalently
\begin{equation}
\Delta\lambda_i=\frac{L}{N},
\qquad i=1,\ldots,N.
\label{eq:supp-open-loop-equal-steps}
\end{equation}
The minimum open-loop work is therefore
\begin{equation}
W_N^{\mathrm{open}}
=\frac{k}{2}\frac{L^2}{N}
=\frac{kL^2}{2N}.
\label{eq:supp-open-loop}
\end{equation}
Using $\mathcal{L}^2=\beta kL^2$, this becomes
$k_{\mathrm B}T\,\mathcal{L}^2/(2N)$, the minimum dissipation of an
$N$-step equilibrated protocol~\cite{nulton1985quasistatic}. It coincides with
the transport term of the closed-loop result~\eqref{eq:supp-min-work-chi}.
The expected work advantage provided by feedback is therefore
\begin{equation}
W_N^{\mathrm{open}}-W_N^*
=
\frac{k_\mathrm{B}T}{2}\,\chi\,\bigl(N-\mathcal{H}_N\bigr),
\label{eq:supp-feedback-advantage}
\end{equation}
which vanishes for an uninformative channel ($\chi=0$) and is maximal under full observation ($\chi=1$).

\section{Thermodynamic-length form of the transport term}
\label{sec:supp-thermo-length}

The thermodynamic length of the equilibrium family is determined by the
Fisher information metric associated with the control parameter $\lambda$.
For the harmonic family,
\begin{equation}
\ln p_\lambda^{\mathrm{eq}}(x)
=-\frac{\beta k}{2}(x-\lambda)^2-\ln Z_\lambda.
\end{equation}
Because translating the trap does not change $Z_\lambda$
(cf.\ \eqref{eq:supp-partition}), the normalization term does not contribute
to the score derivative. Hence
\begin{equation}
\partial_\lambda \ln p_\lambda^{\mathrm{eq}}(x)
=\partial_\lambda\!\left[-\tfrac{\beta k}{2}(x-\lambda)^2\right]
=\beta k(x-\lambda).
\label{eq:supp-score}
\end{equation}
Taking the expectation with respect to $p_\lambda^{\mathrm{eq}}$ gives
\begin{align}
g_{\lambda\lambda}
&:=\mathbb{E}\!\left[
\bigl(\partial_\lambda\ln p_\lambda^{\mathrm{eq}}(x)\bigr)^2
\right]
=\beta^2k^2\,\mathbb{E}[(x-\lambda)^2]
=\beta^2k^2\sigma_{\mathrm{eq}}^2
=\beta k,
\label{eq:supp-fisher}
\end{align}
where the last equality uses
$\sigma_{\mathrm{eq}}^2=(\beta k)^{-1}$. Since the metric is independent of
$\lambda$, the thermodynamic length of the path from $0$ to $L$ is
\begin{equation}
\mathcal{L}
=
\int_0^L \sqrt{g_{\lambda\lambda}}\,d\lambda
=
\sqrt{\beta k}\,L,
\qquad
\mathcal{L}^2=\beta k L^2.
\label{eq:supp-length}
\end{equation}
Equivalently,
$kL^2=\mathcal{L}^2/\beta=k_{\mathrm B}T\,\mathcal{L}^2$.
The transport term in \eqref{eq:supp-min-work-chi} can therefore be written as
\begin{equation}
\frac{kL^2}{2N}
=
\frac{k_\mathrm{B}T}{2}\,\frac{\mathcal{L}^2}{N},
\label{eq:supp-transport-length}
\end{equation}
which recovers the decomposition used in the main text.

\section{Optimal number of feedback cycles under per-cycle overhead}
\label{sec:supp-cadence}

When each feedback cycle carries a fixed overhead cost $c$ (representing measurement, actuation, or fuel expenditure), the total cost of reaching the target in $N$ steps is
\begin{equation}
\mathcal{J}_N = W_N^* + cN
=
\frac{kL^2}{2N}
-\frac{k_{\mathrm B}T}{2}\,\chi\,\bigl(N-\mathcal{H}_N\bigr)
+cN,
\label{eq:supp-total-cost}
\end{equation}
where $W_N^*$ is the partial-observation minimum work from
Eq.~\eqref{eq:supp-min-work-chi}. We seek the minimizing number of
interventions among the admissible integers $N\geq1$.

\subsection{Exact discrete marginal cost}

The marginal cost of adding one additional intervention is
\begin{equation}
\Delta\mathcal{J}_N := \mathcal{J}_{N+1}-\mathcal{J}_N.
\label{eq:supp-DeltaJ-def}
\end{equation}
We can calculate the contributions from the three terms in \eqref{eq:supp-total-cost} independently. \\
The transport term gives
\begin{equation}
\frac{kL^2}{2(N+1)}-\frac{kL^2}{2N}
=
-\frac{kL^2}{2N(N+1)}.
\label{eq:supp-transport-marginal}
\end{equation}
The extraction term gives
\begin{align}
&-\frac{k_{\mathrm B}T}{2}\,\chi
\bigl[(N{+}1-\mathcal{H}_{N+1})-(N-\mathcal{H}_N)\bigr]
\notag\\
&\qquad
=-\frac{k_{\mathrm B}T}{2}\,\chi\!\left(1-\frac{1}{N+1}\right)
=-\frac{k_{\mathrm B}T\chi}{2}\,\frac{N}{N+1},
\label{eq:supp-extraction-marginal}
\end{align}
where we used $\mathcal{H}_{N+1}-\mathcal{H}_N=1/(N{+}1)$. Combining with the overhead increment $c$,
\begin{equation}
\boxed{
\Delta\mathcal{J}_N
=
c
-\frac{kL^2}{2N(N+1)}
-\frac{k_{\mathrm B}T\chi}{2}\,\frac{N}{N+1}.
}
\label{eq:supp-marginal}
\end{equation}
Adding one more cycle is beneficial whenever $\Delta\mathcal{J}_N<0$, i.e.\ whenever the combined transport savings and information-extraction gain exceed the overhead~$c$.

\subsection{Marginal balance and closed-form optimal number of feedback cycles}

Setting $\Delta\mathcal{J}_N=0$ and multiplying through by $2N(N+1)$ yields
\begin{equation}
2cN(N+1)-kL^2-k_{\mathrm B}T\chi\,N^2=0.
\label{eq:supp-marginal-zero}
\end{equation}
Expanding and collecting powers of $N$,
\begin{equation}
(2c-k_{\mathrm B}T\chi)\,N^2+2cN-kL^2=0.
\label{eq:supp-quadratic-raw}
\end{equation}
No large-$N$ or continuum approximation has been made: the discrete marginal condition is exactly quadratic in $N$.

Defining the dimensionless overhead
\begin{equation}
\hat{c}:=\frac{2c}{k_{\mathrm B}T}
\label{eq:supp-chat-def}
\end{equation}
and using the thermodynamic length $\mathcal{L}^2=\beta kL^2$ from Section~\ref{sec:supp-thermo-length}, Eq.~\eqref{eq:supp-quadratic-raw} divides through by $k_{\mathrm B}T$ to give
\begin{equation}
\boxed{
(\hat{c}-\chi)\,N^2+\hat{c}\,N-\mathcal{L}^2=0.
}
\label{eq:supp-quadratic}
\end{equation}

\textbf{Supercritical overhead ($\hat{c}>\chi$)}.
When $\hat{c}>\chi$, i.e.\ $c>\chi\,k_{\mathrm B}T/2$, the per-cycle overhead exceeds the asymptotic per-step information gain. Applying the quadratic formula,
\begin{equation}
\boxed{
N^*
=
\frac{-\hat{c}+\sqrt{\hat{c}^{\,2}+4(\hat{c}-\chi)\,\mathcal{L}^2}}{2(\hat{c}-\chi)}.
}
\label{eq:supp-Nstar}
\end{equation}
This root is the exact marginal-balance point: $\Delta\mathcal J_N<0$ below
it and $\Delta\mathcal J_N>0$ above it. If $N^*$ is not an integer, the
integer optimum is $\max\{1,\lceil N^*\rceil\}$. If $N^*$ is an integer,
$N^*$ and $N^*+1$ have equal cost because
$\Delta\mathcal J_{N^*}=0$.

\textbf{Critical overhead ($\hat{c}=\chi>0$)}.
At the critical value, the quadratic term in
\eqref{eq:supp-quadratic} vanishes and the marginal cost reduces to
\begin{equation}
\Delta\mathcal J_N
=\frac{k_{\mathrm B}T}{2}\,
\frac{\chi N-\mathcal L^2}{N(N+1)}.
\label{eq:supp-critical-marginal}
\end{equation}
It changes sign at the finite balance point
\begin{equation}
N^*\big|_{\hat c=\chi}
=\frac{\mathcal L^2}{\chi}.
\label{eq:supp-Nstar-critical}
\end{equation}
The corresponding integer optimum follows from the same rounding rule as in
the supercritical case.

\textbf{Subcritical overhead ($\hat{c}<\chi$)}.
In this regime the asymptotic information gain per cycle exceeds its
overhead:
\begin{equation}
\lim_{N\to\infty}\Delta\mathcal J_N
=c-\frac{k_{\mathrm B}T\chi}{2}<0.
\end{equation}
The model has no finite optimum in the subcritical regime and favors arbitrarily many feedback cycles.

\end{document}